\documentclass{JHEP3}
\usepackage{graphicx}
\usepackage{epsfig,amsmath,amsthm}

\newcommand{\reef}[1]{(\ref{#1})}
\newcommand{\be}{\begin{equation}}
\newcommand{\ee}{\end{equation}}
\newcommand{\bea}{\begin{eqnarray}}
\newcommand{\eea}{\end{eqnarray}}
\def\bse{\begin{subequations}}
\def\ese{\end{subequations}}

\def\IZ{\relax\ifmmode\hbox{Z\kern-.4em Z}\else{Z\kern-.4em Z}\fi}

\newcommand{\non}{\nonumber \\}

\def\del{{\partial}} \def\dag{\dagger}

\def\al{\alpha}

\def\whG{{\widehat G}}

\def\bi{\begin{itemize}} \def\ei{\end{itemize}}

\def\({\left(} \def\){\right)}
\def\[{\left[} \def\]{\right]}

\title{ \center{On the dynamical generation of the Maxwell term and scale invariance.}}

\author{Eliezer Rabinovici\\
 {\it Racah Institute of Physics, Hebrew University\\
  Jerusalem 91904, Israel} \\
 \email{eliezer@vms.huji.ac.il}
}

\author{Michael Smolkin\\
 {\it Perimeter Institute for Theoretical Physics\\
       Waterloo, Ontario N2L 2Y5, Canada}\\
   \email{msmolkin@perimeterinstitute.ca}
 }

\abstract{Gauge theories with no Maxwell term are investigated in various setups.
The dynamical generation of the Maxwell term is correlated to the scale invariance properties of the system.
This is discussed mainly in the cases where the gauge coupling carries dimensions.
The term is generated when the theory contains a scale explicitly, when it is asymptotically free and in particular also
when the scale invariance is spontaneously broken. The terms are not generated when the scale invariance is maintained.
Examples studied include the large $N$ limit of the $CP^{N-1}$ model in $(2+\epsilon)$ dimensions, a 3D gauged $\phi^6$ vector model and its supersymmetric extension. In the latter case the generation of the Maxwell term at a fixed point is explored. The phase structure of the $d=3$ case is investigated in the presence of a Chern-Simons term as well. In the supersymmetric $\phi^6$ model the emergence of the Maxwell term is accompanied by the dynamical generation of the Chern-Simons term and its multiplet and dynamical breaking of the parity symmetry.
In some of the phases long range forces emerge which may result in logarithmic confinement. These include a dilaton exchange which plays a role also in the case when the theory has no gauge symmetry. Gauged Lagrangian realizations of the 2D coset models do not lead to emergent
Maxwell terms. We discuss a case where the gauge symmetry is anomalous.
}

\begin{document}

%\date{\Huge\today}

\section{Introduction}

The idea of a local gauge symmetry and the experimental discovery of
several gauge particles have played a crucial role in creating and
establishing the standard model. The amazing success of the concept
has not prevented over the years casting doubts on how fundamental the concept is.
These range from the Kaluza-Klein \cite{Kaluza-Klein} theories in which the four dimensional gauge invariance is nothing but the shadow of the five dimensional general covariance, through suggesting that the gauge particles are not themselves elementary objects but only bound states \cite{Bander:1983mg,Seiberg:1994pq,Suzuki:2010yp,Komargodski:2010mc} to pointing out that the local symmetry presentation is just a redundancy. In the language of the early 21st century, gauge symmetries could be emergent.

Among the models realizing these ideas is a representation of $\mathcal{N}=8$
supergravity. There are several recent suggestions of indications
that the theory although seeming superficially non renormalizable is
actually finite \cite{Dixon:2010gz}.  Even though there are cautionary objections to
this claim \cite{Vanhove:2010nf} it would be interesting to speculate on what would be the
possible spectrum of such a finite theory. In particular there were
suggestions that some, but perhaps not all \cite{Bossard:2010dq}, of its gauge symmetries do get
realized by effective propagating gauge bosons the number of which
could be appropriate for the standard model. This should happen at some yet to be determined fixed point or fixed surface in some theories of gravity. In this case it would be satisfying if all scales are dynamically generated.

This leads us to reexamine a class of models in which the gauge
symmetry itself is present a priori, i.e. is not emergent, but the
original Lagrangian does not contain dynamical gauge fields. The
gauge fields have no Maxwell  term to start with. In general
the experience with field theory is that any term which is not
forbidden by some symmetry will emerge in the renormalization
process even if absent in the classical Lagrangian. Thus also the
Maxwell term should emerge in a generic gauge invariant theory.

A symmetry which could enforce the absence of the Maxwell term is
scale/conformal invariance in a system in which the gauge coupling is not
dimensionless. A particular example of that are the so called coset
models of two dimensional Conformal Field Theory(CFT) \cite{Bardakci:1970nb}. Their
Lagrangian realizations \cite{Nahm:1987pc,Bardakci:1987ee,Gawedzki:1988hq} involve
locally gauging some of the global currents of a conformal theory with some
algebraic structure, such as WZW models, without adding a Maxwell term
for the gauge fields. The theory flows from one CFT to another(with
a lower central charge) shedding off its massive confined states, while the conformal symmetry prevents the
Maxwell term from emerging.

In this paper we wish to examine also if the Maxwell term emerges in the
case that there is a scale symmetry but it is spontaneously
broken, such perhaps would be the circumstances if $\mathcal{N}=8$ or some
similar theory turn out to be finite and generate dynamically the scales needed for gravity.

The structure of the paper is to first review in section \ref{sec:d-dim-model} how in the presence of massive dynamical fields the Maxwell term emerges when the gauge coupling carries
a dimension.

In section \ref{sec:CPmodel} we add to the study of the coset models an
examination of the $CP^{N-1}$ model \cite{Eichenherr:1978qa}-\cite{Haber:1980uy}
in $(2+\epsilon)$ dimensions above the region of power counting renormalizability of the model, in particular at its conformal point.

In sections \ref{sec:scalarCS} and \ref{sec:SUSY-CS} we study a gauged version of the $O(N)$ conformal vector model in three dimensions without \cite{Bardeen:1983rv} and with  \cite{Bardeen:1984dx} $\mathcal{N}=1$ supersymmetry (SUSY) respectively. The models exhibit a phase in which the scale symmetry is spontaneously broken. We uncover an extra subtlety in these models: the large $N$ and the IR do not commute. The dilaton plays a special role, note that this massless particle is a singlet under the group and thus the low energy effective Lagrangian consists only of singlets. Some consequences of these facts are discussed. This is studied also in the presence of a Chern-Simons(CS) term, see e.g. \cite{Dunne:1998qy} for a comprehensive review of the subject. While in the nonsupersymmetric case one has the freedom to either introduce the CS term into the action or not, in SUSY it emerges dynamically even if it was absent ab-initio. In particular, the parity symmetry is dynamically broken and the emergence of the Maxwell and CS terms is accompanied by a dynamical generation of their superpartners - the gaugino kinetic and mass terms respectively.

In section \ref{sec:anom2D} we return to the $G/H$ coset models, but this time gauging, what was carefully avoided in the past \cite{Halliday:1985tg}, anomalous subgroups $H$ of $G$. The conclusion section includes a short discussion of the 4D cases in which the gauge coupling is dimensionless.

\section{Dynamical generation of the Yang-Mills term in $d$ dimensions.}
\label{sec:d-dim-model}

In this section we consider a
gauge invariant action of massive complex scalar fields $\Phi$ in a
$d$-dimensional spacetime which does not contain a Yang-Mills term for
the gauge fields. The action is
\begin{equation}
 S\left( \Phi\, , \, \mathcal{A} \right) =
 \int \[\overline{D_{\mu}\Phi} \, D ^{\mu }\Phi
 +m^{\,2}\Phi\overline{\Phi}\]\, d^{d}x \, ,
 \label{scalaraction}
\end{equation}
where $D_{\mu}=\partial_{\mu}+i\mathcal{A}_{\mu}$ is the covariant
derivative and $m$ is the mass of the field $\Phi$. In particular, for
$SU(N_c)$ $\mathcal{A}_{\mu}=A_\mu^a T^a$, $1\leq a \leq N_c^2-1$,
with the generators $\{T^a\}$ normalized by
 \be
 [T^a,T^b]=if^{abc}T^c~ ,\quad
 \texttt{tr}\bigl(T^aT^b\bigr)={1\over 2}\delta_{ab}~ ,
 \ee
where $f^{abc}$ are the structure constants and are antisymmetric in
all indices. In the presence of only a covariant derivative the coupling constant could have been and was absorbed   in the definition of the gauge field. Quantum loop corrections generate the Yang-Mills term for the gauge field $\mathcal{A}_{\mu}$
when the theory is superrenormilizable. The gauge coupling turns out
to be proportional to the mass of the scalar field raised to the
power determined by dimensional analysis.

Integrating out the $\Phi$ degrees of freedom leads to a
non-local effective action for the gauge field $A_{\mu}$
 \be
 e^{-S_{eff}(\mathcal{A})}=\int D\Phi \, D\overline{\Phi} e^{-S\left(\Phi , \mathcal{A} \right)} ~~ ,
 \ee
where
 \be
 S_{eff}(\mathcal{A})= \texttt{Tr}\ln \left( -D_{\mu}D^{\mu} +m^{\,2} \right) ~ .
 \label{eqn:Aeff-action}
 \ee

Expanding $S_{eff}(\mathcal{A})$ about $ \mathcal{A}_{\mu}=0$ and keeping
only quadratic terms in the field gives\footnote{Throughout this
section lower case $``\texttt{tr}"$ represents trace over color
indices.} (see Appendix \ref{appx:bfncl-det-expan} for details of
the derivation)
 \begin{multline}
 S_{eff}(\mathcal{A})=
 {\Gamma(2-d/2) \over 6(4\pi)^{d/2}}~\texttt{tr}\int {d^{d} p \over (2\pi)^d }  \mathcal{A}_{\mu}(p) \mathcal{A}_{\nu}(-p)
 \[ \delta^{\mu\nu} - {p^{\,\mu}p^{\,\nu} \over p^{\,2}} \] \, p^{\,d-2}
 \\ \times \({1\over 4}+{m^2\over p^{\,2} }\)^{d/2-2} \ _{2}F_1\({3\over 2} ~,~ {4-d \over 2} ~,~{5\over 2} ~;~ {1\over 1+4 \, m^2/p^2}\)
 \, + \, \mathcal{O}\(\mathcal{A}^3\) \, .
 \label{eqn:quad-Aeff-action}
 \end{multline}
 In particular, choosing the Lorentz gauge fixing term
\be
 S_{GF}=m^{d-4}\, {\Gamma(2-d/2) \over 6(4\pi)^{d/2}}\int d^{d} x\(\partial_{\mu}A^{\mu}\)^2
 = m^{d-4}\,{\Gamma(2-d/2) \over 6(4\pi)^{d/2}}\int_p p_{\mu}p_{\nu}A^{\mu}(p)A^{\nu}(-p) \,,
 \ee
leads to the following propagator for the gauge field
 \be
  \langle \mathcal{A}_\mu^a(p)\mathcal{A}_{\nu}^b(-p)\rangle = { 6(4\pi)^{d/2} \over \Gamma(2-d/2)} \,
  {  m^{4-d}  \over p^2\, \gamma(p)}\,\delta^{ab}
  \( \delta_{\mu\nu}-{p_{\mu}p_{\nu}\over p^2}\[ 1-\gamma(p)\] \)~,
 \ee
where
 \be
 \gamma(p) \equiv
 \, \(1+{p^{\,2}\over 4\,m^2}\)^{d/2-2} \ _{2}F_1\({3\over 2} ~,~ {4-d \over 2} ~,~{5\over 2} ~;~ {1\over 1+4 \, m^2/p^2}\)\, .
 \ee

In the limit $p<<m$, we obtain
 \begin{multline}
 S_{eff}(\mathcal{A})= m^{d-4} \, {\Gamma(2-d/2) \over 6(4\pi)^{d/2}}~\texttt{tr}\int {d^d p \over
 (2\pi)^d }  \mathcal{A}_{\mu}(p) \mathcal{A}_{\nu}(-p)
 \[ \delta^{\mu\nu}p^{\,2} - p^{\mu}p^{\nu} \] + \, . \, . \, .
 \\ = \,m^{d-4} ~ {\Gamma(2-d/2) \over 12(4\pi)^{d/2}} \int d^d x
 ~{1\over 2}\bigl(F_{\mu\nu}^a\bigr)^2 + \, . \, . \, .
 \label{eqn:low-energy-Aeff-action}
 \end{multline}
with $F_{\mu\nu}^a=\del_{\mu}A_{\nu}^a-\del_{\nu}A_{\mu}^a
-f^{abc}A_{\mu}^b A_{\nu}^c$. In the last equality we used gauge
invariance of the action to supplement the quadratic part with
appropriate self-interaction terms of $A_{\mu}^a$ that make gauge
invariance explicit. Although $A_{\mu}^a$ were originally introduced
as dummy fields a Yang-Mills term is dynamically generated for
all of them. The fact that the Yang-Mills term is generated for all
$A_{\mu}^a$ fields rather than for a smaller subset results from the
gauge invariance of the action, and the gauge coupling scales
appropriately with $m$.

A comment should be made regarding the situation when the action
depends on a collection of distinct complex scalar fields charged
under the gauge group. In this case, according to the above
discussion each massive field contributes to the Yang-Mills
term. In particular, if an $SU(N_f)$ vector multiplet $\vec{\Phi}$
is introduced into the action, then the coupling charge decreases as
$1/\sqrt{N_f}$ and becomes arbitrarily small in the large $N_f$
limit.

In the ultrarelativistic limit $p>>m$ or equivalently in the
massless case $m=0$, we obtain
\begin{multline}
 S_{eff}(\mathcal{A})= -\, {\Gamma(1-d/2) \Gamma[d/2]\over 2 (16\pi)^{d-1 \over 2}\Gamma[d/2+1/2]}
 ~\texttt{tr}\int {d^d p \over
 (2\pi)^d }  \mathcal{A}_{\mu}(p) \mathcal{A}_{\nu}(-p)
 \[ \delta^{\mu\nu}p^{\,2} - p^{\mu}p^{\nu} \] p^{\,d-4} + \, . \, . \, .
 \\ = {\sqrt{\pi}\over 16(2\pi)^{\,d}} ~ {\Gamma(d/2-1)\Gamma(d-2) \over \Gamma[d/2+1/2]}
 \int d^dx d^dy ~ {\(F^a\)^{\mu\nu}(x) \(F^a\)_{\mu\nu}(y) \over 2|x-y|^{2d-4}} + \, . \, . \, .
 \label{eqn:low-energy-Aeff-action2}
\end{multline}

Hence, the theory explicitly manifests a non-local character. However, this conclusion is a reflection of the ultrarelativistic limit. In this case we probe the short wavelength physics. The gauge particles are introduced into the action as dummy variables and eventually are built out of the scalar $\Phi$ particle degrees of freedom. As a result, one indeed might expect that the short distance behavior of the gauge field will reveal a non-local structure.

For those cases in which the gauge coupling is a dimensionful parameter the Yang-Mills term was generated.
This suggests that once there is no scale in the problem or all the scales
at hand are irrelevant, the Yang-Mills term will not be generated.

More generally, if the theory possesses a symmetry, for example a scale/conformal symmetry, which forbids any scale
generation then it also excludes dynamical generation of the
long-range force mediated by the gauge particles originally
introduced into the action as dummy fields.  At least as long as a symmetry is not spontaneously broken.

In what follows, we demonstrate how this general argument manifests
itself in the case of large $N$ vector models with a $U(1)$ gauge symmetry.

\section{The $\mathbf{CP}^{N-1}$ model in $(2+\epsilon)$-dimensions.}
\label{sec:CPmodel}

In this section we discuss the non emergence of the Maxwell term at a fixed point.
This behavior is exhibited already in two dimensions in the
so called coset models \cite{Bardakci:1987ee,Nahm:1987pc,Gawedzki:1988hq} mentioned in the
introduction. Here we study the $CP^{N-1}$ model in
$(2+\epsilon)$ dimensions at its fixed point \cite{Duane:1979in}. While in two
dimensions this asymptotically free model served as an example how
the dynamical mass generation leads to a long range force, that is a
Maxwell term for a gauge field, in $(2+\epsilon)$ dimensions the
model has a fixed point at which it is scale invariant \cite{Polyakov:1975rr}. This fixed point can be reliably analyzed for large $N$ using the $\epsilon$ expansion.

We review in section \ref{subsec:CPgap} and \ref{sec:CPn-critical-point} the essential properties of the model starting by recalling the gap equations and showing that their solutions reveal three possible phases of the system - the weakly
coupled phase, where the global $SU(N)$ symmetry is spontaneously broken and the $U(1)$ gauge potential is short-ranged; the phase in which the $SU(N)$ symmetry is unbroken and the $U(1)$ gauge field confines; and the third phase which is a scale invariant theory at the fixed point separating the two other phases. We describe the renormalization of a four-point function and reaffirm the existence of a fixed point \cite{Duane:1979in}.

In subsection \ref{sec:CPn-Maxwell-gen} we further analyze
the massive phase of the system. It is shown that in the low energy
limit a Maxwell term for the auxiliary field $A_{\mu}$ emerges.
However, it disappears from the action when the coupling constant is
tuned to its critical value.

The Lagrangian as written in terms of constrained fields is
 \be
 \mathcal{L}=\del_{\mu} \vec{z}^{\,\dag} \del^{\mu} \vec{z}+{g_0 \over 4 N}
 ( \vec{z}^{\,\dag} \, \overset{\leftrightarrow}\del_{\mu} \, \vec{z} )^2
 ~ ,
 \label{eqn:CPnLagran}
 \ee
with $\vec{z}$ being an $N$- component column vector satisfying the
following constraint
 \be
 \vec{z}^{\,\dag} \cdot \vec{z} = {N \over g_0} ~ .
 \label{eqn:CPcon}
 \ee
This system exhibits invariance under local gauge transformations
$\vec{z}\rightarrow e^{i\alpha(x)}\vec{z}$. Following \cite{Haber:1980uy} we use it to fix $z_N$ to be real for all spacetime points $x$. 

In order to re-write the Lagrangian in terms of unconstrained fields it is necessary to go to a non-linear representation of the coset
space \cite{Eichenherr:1978qa}
 \be \mathrm{CP}^{N-1}\simeq \mathrm{SU}(N)/\mathrm{S}(\mathrm{U}(1)\times \mathrm{U}(N-1)) \, .\label{eqn:coset-patt}\ee
Rather than presenting the details of such a transformation we
address the reader to \cite{Bardeen:1976zh,Haber:1980uy} and
references therein. The action of the $CP^{N-1}$ model in terms of
the unconstrained variables is given by
 \be
 S\left( \vec{\phi}\, , \, \vec{\phi}^{\,\dag}\right) =\int d^{\,2+\epsilon}x
 \[{\del_{\mu}\phi_i^{\,\dag} \, \del ^{\mu}\phi_i \over
 \(1+\frac{g_0}{4N} \phi_i \phi_i^{\dag}\)^{2}}
 +\frac{g_0}{4N}{ \(\phi_i^{\dag} \, \overset{\leftrightarrow}\del_{\mu} \, \phi_i\)^2 \over
 \(1+\frac{g_0}{4N} \phi_i \phi_i^{\dag}\)^{4}} \] ~ ,
 \ee
where $i=1,..,N-1$ and
 \be
 z_i={\phi_i \over 1+ {g_0 \over 4 N} \vec{\phi}^{\,\dag} \cdot \vec{\phi} }
 \quad .
 \ee
The real $z_N$ field is determined through constraint \reef{eqn:CPcon}. Note that the original gauge invariance of the model
is only partially fixed. For configurations in which the $N$'s component of $\vec{z}$ is non vanishing the gauge invariance is indeed fixed.
However, when it vanishes, i.e. on a subset of the constrained $\vec{\phi}$ fields which obey
 \be
 \vec{\phi}^{\,\dag} \cdot \vec{\phi} = {4 N \over g_0} ~ ,
 \label{eqn:GribovSet}
 \ee
there remains a residual gauge symmetry \cite{Haber:1980uy},
by analogy referred as a Gribov ambiguity. The choice of having the $z_N$ component real constitutes a gauge fixing only as long as this component is non vanishing. When it does vanish the gauge choice is performed by choosing the first non vanishing component of $\vec z$ down the line to be real. For all these choices there are Gribov ambiguities which lead to the same physics result. A similar statement can be made when the gauge fixing is centered around attempting to choose any other component, $z_i$, as real first.

As will be shown below, in the large $N$ limit one out of three
accessible phases of the system dynamically emphasizes this subset and it is
accompanied by an appearance of the Maxwell term for the U(1) gauge field.
In the other phases the Gribov ambiguity plays no important role.

\subsection{Generating functional and the gap equations.}
\label{subsec:CPgap}

Rescaling the fields according to
 \be
 \vec{\phi}=Z_{\phi}^{1/2} \vec{\phi}_r \, ,
 \ee
where $Z_{\phi}$ is an arbitrary real constant, we obtain
 \be
 S\left( \vec{\phi}_r\, , \, \vec{\phi}^{\,\dag}_r \right) =  \int d^{\,2+\epsilon}x
 \[
 {Z_{\phi} \, \del_{\mu}\vec{\phi}_r^{\,\dag} \, \del ^{\mu}\vec{\phi}_{r} \over
 \(1+\frac{Z_{\phi}}{4N}g_0 \vec{\phi}_{r} \vec{\phi}_{r}^{\dag}\)^{2}}
 +\frac{g_0}{4N}Z_{\phi}^2{ \(\vec{\phi}_{r}^{\,\dag} \, \overset{\leftrightarrow}\del_{\mu} \, \vec{\phi}_{r}\)^2 \over
 \(1+\frac{Z_{\phi}}{4N}g_0 \vec{\phi}_{r} \vec{\phi}_{r}^{\dag}\)^{4}}
 \]  ~ .
 \ee
Introducing the following identities
\begin{equation}
 1\sim\int D\eta_{\mu} \, \delta \(i \, \vec{\phi}_{r}^{\,\dag} \, \overset{\leftrightarrow}\del_{\mu} \, \vec{\phi}_{r}+ N \eta_{\mu}\)
 \sim \int D A_{\mu} \, D\eta^{\mu} \, e^{i\int A^{\mu}
 \(N \eta_{\mu} + i \, \vec{\phi}_{r}^{\,\dag} \, \overset{\leftrightarrow}\del_{\mu} \, \vec{\phi}_{r}\)d^{\,2+\epsilon}x}~,
\label{eqn:identity-2}
\end{equation}
and
\begin{equation}
 1\sim\int D\rho \, \delta(\vec{\phi}_r\cdot\vec{\phi}_r^{\,\dag}-N\rho)
 \sim \int D\rho \, D\lambda \, e^{-i\int \lambda( \vec{\phi}_r\cdot\vec{\phi}_r^{\,\dag}-N\rho )d^{3}x}
\label{eqn:identity}~,
\end{equation}
into the generating functional for $\vec{\phi}_r$ yields\footnote{Index $r$ is suppressed for brevity and $N\sim
N-1$ is used.}
\begin{eqnarray}
 Z\left[  \vec{J}, \, \vec{J}^{\,\dag} \right]&=& \int D\rho D\lambda\,DA_{\mu}D\eta^{\mu}
 \, e^{- S_{eff}(\rho,\,\lambda, \, A,\eta) }
 \non
  &&\quad\quad\quad\times e^{\int\vec{J}^{\,\dag}(x)\left[ -Z_{\phi}\del_{\mu}
 \( 1+\frac{Z_{\phi}}{4}g_0 \rho \)^{-2}\del^{\mu} +i\lambda
 +A^{\mu}\overset{\leftrightarrow}\del_{\mu}\right]_{xy}^{-1}\vec{J}(y)~d^{\,2+\epsilon}x~d^{\,2+\epsilon}y}
 \, ,
 \label{eqn:CPNpartition}
\end{eqnarray}
where
\begin{eqnarray}
 S_{eff}(\rho,\,\lambda, \, A,\eta)=-N&\int& d^{\,2+\epsilon}x \, \[ i\lambda \, \rho
 +i\,\eta_{\mu}A^{\mu}+\frac{g_0}{4}Z_{\phi}^2{ \eta_{\mu}\eta^{\mu} \over
 \(1+\frac{Z_{\phi}}{4}g_0 \rho\)^{4}}\]
 \non
 &+& N\texttt{Tr}\ln \left[ -Z_{\phi}\del_{\mu}\( 1+\frac{Z_{\phi}}{4}g_0 \rho \)^{-2}\del^{\mu} +i\lambda
 +A^{\mu}\overset{\leftrightarrow}\del_{\mu}\right]  \, .
\end{eqnarray}

The above action is both quadratic in the auxiliary field
$\eta_{\mu}$ and does not contain its derivatives. Hence,
$\eta_{\mu}$ can be eliminated by using its equations of
motion, and one arrives at
\begin{eqnarray}
 S_{eff}(\rho,\,\lambda, \, A)=-N&\int& d^{\,2+\epsilon}x \, \[ i\lambda \, \rho
 +{ \(1+\frac{Z_{\phi}}{4}g_0 \rho\)^{4} \over g_0 Z_{\phi}^2}A_{\mu}A^{\mu}\]
 \non
 \quad\quad
 &+& N\texttt{Tr}\ln \left[ -Z_{\phi}\del_{\mu}\( 1+\frac{Z_{\phi}}{4}g_0 \rho \)^{-2}\del^{\mu} +i\lambda
 +A^{\mu}\overset{\leftrightarrow}\del_{\mu}\right]  \, .
 \label{eqn:CPNSeff}
\end{eqnarray}

In the next section we show that the vector field $A_{\mu}$ plays a
role of the $U(1)$ gauge field as a consequence of the Gribov ambiguity
mentioned earlier.

The Lorentz invariant gap equations are given for the large $N$ by
the stationary phase approximation\footnote{Note that the original gauge symmetry of the model was fixed up to a possible Gribov ambiguity.}
 \bea
 &&\langle A_{\mu}\rangle =0 ~,
 \non
 &&i\bar{\lambda}=g_0\frac{Z_{\phi}^{\,2}}{2}\( 1+\frac{Z_{\phi}}{4}g_0 \bar{\rho} \)^{-3}
 \texttt{tr}\[ \frac{\del^2}
 {-Z_{\phi}\( 1+\frac{Z_{\phi}}{4}g_0 \bar{\rho} \)^{-2}\del^2 +i\bar{\lambda}} \] ~,
 \non
 &&\bar{\rho}=\texttt{tr}\[ \frac{1}
 {-Z_{\phi}\( 1+\frac{Z_{\phi}}{4}g_0 \bar{\rho} \)^{-2}\del^2 +i\bar{\lambda}}\] ~.
 \eea
Choosing
 \be
 Z_{\phi}^{1/2}=\frac{2}{1+(1-g_0 \bar{\rho})^{1/2}} \, ,
 \ee
in order to fix the residue of the scalar field propagator to $1$,
simplifies the gap equations as follows
 \bea
 m^2&=&-g_0\frac{Z_{\phi}^{\,1/2}}{2}
 \texttt{tr}\[ \frac{-\del^2}
 {-\del^2 +m^2} \]
 = m^2 g_0\frac{Z_{\phi}^{\,1/2}}{2}\bar{\rho}~,
 \non
 \bar{\rho}&=&\texttt{tr}\[ \frac{1}
 {-\del^2 +m^2}\] = {\Gamma(-\epsilon/2) \over (4\pi)^{1+\epsilon/2}}(m^2)^{\epsilon/2}~,
 \label{eqn:CPNgap-eqn}
 \eea
where $m^2=i\bar{\lambda}$ stands for the mass of the scalar field,
and the dimensional regularization has been used to evaluate the
divergent loop integral.

From the first equation we recover two out of three possible phases
of the system: the weak coupling and the strong coupling ones.
These phases were studied in \cite{Bardeen:1976zh} for the nonlinear $\sigma$ model
and in \cite{Haber:1980uy,Duane:1979in} for $CP^{N-1}$
 \bea
 &&\text{Phase 1:}\qquad m^2=0~, ~\bar{\rho}=0~, ~Z_{\phi}=1~,
 \non
 &&\text{Phase 2:}\qquad m^2=\[{(4\pi)^{1+\epsilon/2}\over g_0 \Gamma(-\epsilon/2)}\]^{2/\epsilon},
 ~\bar{\rho}={2 \over g_0 Z_{\phi}^{1/2}}
 ={1 \over g_0}~,~ Z_{\phi}=4~.
 \label{eqn:CPNphases}
 \eea

The third phase is derived and explored in the next section.
It corresponds to a scale invariant fixed point that separates between
the phases.

\subsection{$\phi\phi\rightarrow\phi\phi$ scattering amplitude and RG
flow.} \label{sec:CPn-critical-point}

It will be useful to explicitly obtain the various propagators as well as a four-point proper vertex. For this purpose we use
the generating functional  (\ref{eqn:CPNpartition}) and
differentiate it with respect to the sources $\vec{J}$ and
$\vec{J}^{\dag}$
 \be
 \langle\phi_i(x_1)\phi^{\dag}_j(x_2)\phi_k(x_3)\phi_l^{\dag}(x_4)\rangle
 \simeq {\delta^4 \over \delta J^{\dag}_i(x_1)\delta J_j(x_2)\delta J^{\dag}_k(x_3)\delta J_l(x_4)}
 Z\left[  \vec{J}, \, \vec{J}^{\,\dag}\right]\Bigg|_{\vec{J}=\vec{J}^{\dag}=0}
  ~ .
 \ee

\begin{figure}[t!]
\centering \noindent
\includegraphics[width=9cm]{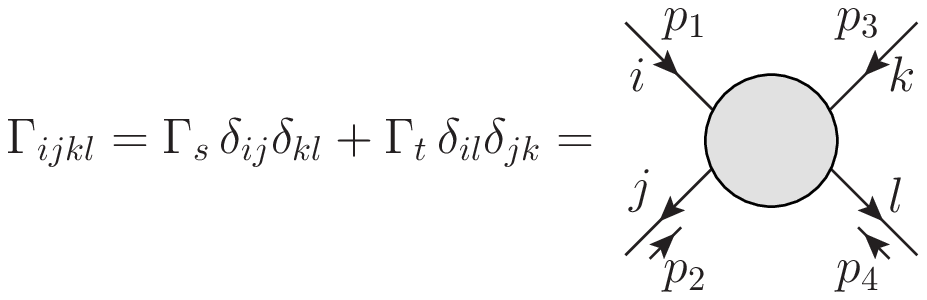}
\caption[]{Diagramatic reperesentation of the full connected
one-particle irreducible amputated four-point function.}
\label{fig:proper-vertex}
\end{figure}

Henceforth we shall deal with $\delta_{ij}\delta_{kl}$ part of the
fully connected one-particle irreducible amputated four-point
function shown on figure \ref{fig:proper-vertex}.
This part is associated with the $s-$channel and applying
appropriate changes ($s\rightarrow t$, $\delta_{ij}\delta_{kl}
\rightarrow \delta_{il}\delta_{jk}$ and $p_1\leftrightarrow p_3$)
one gets similar results for the $t-$channel.

In what follows we omit the various delta-functions, redefine
$i\lambda\rightarrow \lambda$ as well as adopt the notation of
Appendix \ref{appx:bfncl-det-expan}
 \bea
 \int_{x}&\equiv&\int d^{\,2+\epsilon}x \quad,\quad
 \int_{p}\equiv\int {d^{\,2+\epsilon}p \over (2\pi)^d}
 \quad ,
 \non
 \whG&\equiv&(-\del^2+m^2)^{-1} ~ \Leftrightarrow ~
 G(x-y)=\langle x|\whG|y\rangle=\int_{p} {e^{ip(x-y)} \over p^2+m^2}
 \quad .
 \eea
where $x$ represents the spacetime coordinate and $p$ is the
momentum $(2+\epsilon)$-vector.

Carrying out the differentiation with respect to the sources one gets a
pair of propagators inside the path integral. Each such propagator
has the following form
\begin{multline}
 \langle x_m| \left[ -Z_{\phi}\del_{\mu}
 \( 1+\frac{Z_{\phi}}{4}g_0 \rho \)^{-2}\del^{\mu} +\lambda
 +A^{\mu}\overset{\leftrightarrow}\del_{\mu}\right]^{-1} |x_n\rangle
 = G(x_m-x_n)
 \\
 +{g_0\sqrt{Z_{\phi}} \over 2}\delta^{\mu\nu}\int_z {\del \over \del x^{\mu}_m}
 G(x_m-z)\rho(z){\del \over \del x^{\nu}_n}G(x_n-z)
 -\int_z G(x_m-z)\lambda(z)G(x_n-z)
 \\
 +\int_z G(x_m-z)A^{\mu}(z){\del \over \del x^{\mu}_n}G(x_n-z)
 -\int_z {\del \over \del x^{\mu}_m}G(x_m-z)A^{\mu}(z)G(x_n-z)
 +\ldots ~,
 \label{eqn:prop-expansion}
\end{multline}
where $(m,n)$ equals either $(1,2)$ or $(3,4)$ and we have expanded
around the solutions of the gap equations (\ref{eqn:CPNgap-eqn}):
$\rho\rightarrow\bar\rho+\rho,\lambda\rightarrow\bar\lambda+\lambda$,
keeping only linear terms in the small perturbations
$\rho,\lambda\,$.

It turns out, that these terms are sufficient to compute any
correlation function to leading order in the $1/N$ expansion.
Indeed, terms beyond linear order will require to introduce extra
propagators of the auxiliary fields in the computation of a given
correlation function. However, as we shall see below, each such
propagator is inversely proportional to $N$ and therefore the
contribution of higher order terms in the above expansion will be
suppressed by a power of $1/N$ relative to the result based
on the linear terms only.

Accordingly, in order to compute the four-point function it is
enough to evaluate the propagators of the auxiliary fields
$\rho,\lambda$ and $A_{\mu}$. For this purpose we expand the
effective action (\ref{eqn:CPNSeff}) around the solutions $\bar\rho,
\bar\lambda$ of the gap equations (\ref{eqn:CPNgap-eqn}) and keep
only quadratic part in the perturbations. The $\rho-A_{\mu}$ and
$\lambda-A_{\mu}$ mixing terms turn out to vanish. Hence, $A_{\mu}$
is decoupled from the $\rho-\lambda$ sector. Using the gap equations,
one finds that linear terms vanish as well.

According to the results described in Appendix \ref{appx:bfncl-det-expan} we
get the following expression for the term quadratic in $A_{\mu}$ in
the effective action
\begin{multline}
 \mathcal{O}(A^2) = ~ N\,\(\bar\rho-{1 \over g_0}\)
 \int_p ~ A_{\mu}(p) A^{\mu}(-p)
 \\
 +{\,\epsilon\,N\,\bar\rho\over 12}\int_p  A_{\mu}(p) A_{\nu}(-p)
 \[ \delta^{\mu\nu} - {p^{\,\mu}p^{\,\nu} \over p^{\,2}} \] \,\({p \over m}\)^{\epsilon}
 \\ \times \({1\over 4}+{m^2\over p^{\,2} }\)^{\epsilon/2-1}
 \ _{2}F_1\({3\over 2} ~,~ {2-\epsilon\over 2} ~,~{5\over 2} ~;~ {1\over 1+4 \,
 m^2/p^2}\)~.
 \label{eqn:CPn-Aeff-action}
\end{multline}
Adding altogether the quadratic terms in the action leads to
\begin{multline}
 S_{eff}(\rho,\lambda,A)={N\over 2}\int_p
 \bigl[\rho(p)\rho(-p)T_{\rho\rho}(p)+2\lambda(p)\rho(-p)T_{\rho\lambda}(p)+\lambda(p)\lambda(-p)T_{\lambda\lambda}(p)\bigr]
 \\
 + N\,\(\bar\rho-{1 \over g_0}\)\int_p ~ A_{\mu}(p) A^{\mu}(-p)
 +{N\,\epsilon\,\bar\rho\over 12}\int_p  A_{\mu}(p) A_{\nu}(-p)
 \[ \delta^{\mu\nu} - {p^{\,\mu}p^{\,\nu} \over p^{\,2}} \] \,\({p \over m}\)^{\epsilon}
 \\ \times \({1\over 4}+{m^2\over p^{\,2} }\)^{\epsilon/2-1}
 \ _{2}F_1\({3\over 2} ~,~ {2-\epsilon\over 2} ~,~{5\over 2} ~;~ {1\over 1+4 \, m^2/p^2}\)
 +\ldots
 \label{eqn:CPNquadr-action}
\end{multline}
where
 \bea
 T_{\rho\rho}(p)&=&-{g_0^2 Z_{\phi} \over 4} \( p^{\mu}p^{\nu}\,I_{\mu\nu}(p)-2m^2 p^{\mu}I_{\mu}(p)-p^2\bar\rho+m^4I_0(p)-{1\over2}m^2\bar\rho \)
 ~,\non
 T_{\rho\lambda}(p)&=& {g_0 \sqrt{Z_{\phi}} \over 2}\( p^{\mu}I_{\mu}(p)- m^2 I_0(p)+\bar\rho\)-1
 ~,\non
 T_{\lambda\lambda}(p)&=&-I_0(p) ~.
 \eea
with
\begin{multline}
 I_0(p^2)\equiv\int_q \frac{1}{ [q^{\,2}+m^2] [(q+p)^{\,2}+m^2]}
 \\={\Gamma( 1-\epsilon/2) \over (4\pi)^{1+\epsilon/2}}
 \({p^2\over 4}+m^2\)^{\epsilon/2-1}
 \ _{2}F_1\({1\over 2} ~,~ {2-\epsilon \over 2} ~,~{3\over 2} ~;~ {1\over 1+4\, m^2/p^2}\)
 \, , \label{I-0}
\end{multline}
\begin{equation}
 I_{\mu}(p)\equiv\int_q \frac{q_{\mu}}{ [q^{\,2}+m^2] [(q+p)^{\,2}+m^2]}=
 -{1 \over 2}\,I_0(p)\,p_{\mu}
 \quad
 ,\quad\quad\quad\quad\quad\quad\quad\quad\quad\quad\quad\quad\quad\quad~
 \label{I-mu}
\end{equation}
and
 \bea
 I_{\mu\nu}(p)&\equiv&\int_q \frac{q_{\mu}q_{\nu}}{ [q^{\,2}+m^2] [(q+p)^{\,2}+m^2]}
 \non
 &=&\delta_{\mu\nu}{\Gamma( -\epsilon/2) \over 2(4\pi)^{1+\epsilon/2}}
 \({p^2\over 4}+m^2\)^{\epsilon/2}
 \ _{2}F_1\({1\over 2} ~,~ {-\epsilon \over 2} ~,~{3\over 2} ~;~ {1\over 1+4\, m^2/p^2}\)
 \non
 && \, + \, p_{\mu}p_{\nu}{\Gamma( 1-\epsilon/2) \over 4(4\pi)^{1+\epsilon/2}}\({p^2\over 4}+m^2\)^{\epsilon/2-1}
 \non
 &&\times \[  _{2}F_1\({1\over 2} ~,~ {2-\epsilon \over 2} ~,~{3\over 2} ~;~ {1\over 1+4\, m^2/p^2}\)
 +{1\over 3}\ _{2}F_1\({3\over 2} ~,~ {2-\epsilon \over 2} ~,~{5\over 2} ~;~ {1\over 1+4\, m^2/p^2}\)\]
 \, .\non\non
 \label{I-munu}
\eea Note also that based on the above definitions the following
useful identity holds
 \be
 p^{\mu}p^{\nu}I_{\mu\nu}(p)= {p^4 \over 4}I_0(p^2)+{p^2 \over 2}\bar\rho
 ~.
 \ee

Since $\rho-\lambda$ sector is decoupled from $A_{\mu}$ we compute
their contributions to the four-point proper vertex separately. We
start from $\rho-\lambda$. Inverting the matrix $T$, we obtain
\begin{multline}
 D_{\rho\rho}^{(1)}(p)={I_0 \over N(1+{g_0\over 2}p^2I_0)}
 ~ ; ~
 D_{\lambda\lambda}^{(1)}(p)={g_0^2\over 16}\,{p^4I_0 \over N(1+{g_0\over 2}p^2I_0)}
 ~ ; ~
 D_{\lambda\rho}^{(1)}(p)=-{1+{g_0\over 4}p^2I_0 \over N(1+{g_0\over 2}p^2I_0)}~,
 \label{eqn:massless-prop}
\end{multline}
and
\begin{multline}
 D_{\rho\rho}^{(2)}(p)={2 \over N\,g_0} \, {1 \over p^2 + m^2} ~ ;~
 D_{\lambda\lambda}^{(2)}(p)= {g_0 \over 2N} \, {\,p^4 \over p^2+m^2}+{2g_0m^2\over N}-{1 \over N I_0} ~;~
 \\ D_{\lambda\rho}^{(2)}(p) = -{1\over N}-{m^2 \over N(p^2 + m^2)} ~,
 \label{eqn:mass-prop}
\end{multline}
for the Higgs and confinement phases respectively. The upper index
indicates the phase number, whereas lower indices indicate what kind
of propagator is concerned.

The feature of (\ref{eqn:mass-prop}) is that to leading order in
$1/N$ the $\rho$-propagator consists only of a simple pole when rotated
back to the Minkowski space. That is $\vec\phi\vec\phi^{\dag}$
creates a single-particle state when acting on the vacuum. This
particle is degenerate in mass with the $\phi$'s and gets bound by the confining potential into the singlet and the adjoint representations of $SU(N)$ \cite{Haber:1980uy}. This corresponds to a full  restoration of the global $SU(N)$ symmetry in the confined phase.

Let us next use (\ref{eqn:prop-expansion}) and the propagators of
the auxiliary fields to calculate the contribution of the
$\rho-\lambda$ sector to the $s-$channel of the connected
one-particle irreducible amputated four-point function
$\Gamma_{s}^{(r)}$
 \be
 \Gamma_{s}^{(r)}\Big|_{\rho\lambda}={g_0^2 Z_{\phi}\over 4}(p_1\cdot p_2)(p_3\cdot p_4)D_{\rho\rho}^{(r)}(s)
 +D_{\lambda\lambda}^{(r)}(s)+{g_0\sqrt{Z_{\phi}}\over 2}(p_1\cdot p_2+p_3\cdot
 p_4)D_{\lambda\rho}^{(r)}(s) ~,
 \ee
where $r=1,2$ serves to indicate the different phases and the vertical bar with
a subscript $``\rho\lambda"$ indicates restriction of the full one-particle irreducible
amputated four-point function $\Gamma_{s}^{(r)}$ to the $\rho-\lambda$ sector. Substituting
(\ref{eqn:massless-prop}),(\ref{eqn:mass-prop}) yields
 \be
  \Gamma_{s}^{(1)}\(p_i^2=0\)\Big|_{\rho\lambda}= {-g_0 s \over N(2+g_0 s I_0(s))} \quad,
  \label{eqn:scalar-prop-vertex1}
 \ee
for the on-shell vertex in the Higgs phase and
\begin{multline}
 \Gamma_{s}^{(2)}\Big|_{\rho\lambda}= {g_0 \over 2N}(s-m^2-p_1^2-p_2^2){1\over s+m^2}(s-m^2-p_3^2-p_4^2)
\\ +{g_0 \over 2N}\(-s+3m^2+\sum_{i=1}^4 p_i^2\)-{1 \over N I_0(s)}~,
 \label{eqn:scalar-prop-vertex2}
\end{multline}
for the off-shell vertex in the confinement phase.

Let us complete the above expression by adding the contribution from
the exchange auxiliary field $A_{\mu}$. For this purpose we need to compute
its propagator in the two accessible phases separately. In
the Higgs phase according to (\ref{eqn:CPNphases}) and
(\ref{eqn:CPn-Aeff-action}), we obtain
 \be
 D_{\mu\nu}^{(1)}(p)=-{ g_0 \over 2N }\( 1+g_0f^{(1)}p^{\,\epsilon} \)^{-1}\( \delta_{\mu\nu}+ g_0f^{(1)}p^{\,\epsilon} \, {p_{\mu}p_{\nu} \over p^2}\)
 ~,
 \label{prop1}
 \ee
with
 \be
 f^{(1)}={-\epsilon  \over 4^{2+\epsilon} ~ \pi^{\epsilon/2+1/2}}
 {\Gamma(-\epsilon/2)\Gamma(\epsilon/2) \over \Gamma(3/2+\epsilon/2)}
 ={1 \over (1+\epsilon)g_c}
 \quad,
 \ee
where
 \be
 g_c=2(4\pi)^{1+\epsilon/2}{ \Gamma(\epsilon) \over \Gamma\( 1-{1\over 2}\epsilon \) \Gamma(\epsilon/2)^2}
 \quad ,
 \ee
and we show below that $g_c$ is an ultraviolet-stable fixed point of the
renormalization group flow. The propagator of the auxiliary field (\ref{prop1}) has no pole at $p=0$. Thus we denote this phase as a Higgs phase.

However, in the confinement phase $A_{\mu}$ behaves like a gauge vector field
and it will be necessary to further fix the gauge in order to proceed. The need to further fix the gauge is tightly related to the Gribov ambiguity mentioned earlier. In fact, it occurs because the vacuum has chosen to localize on the ambiguous field configurations.  Indeed, in the confinement phase according to (\ref{eqn:CPNphases}), we have
 \be
 \langle \, \vec{\phi}^{\,\dag}_r \cdot \vec{\phi}_r \,\rangle
 =N\bar\rho={N \over g_0}
  \quad \quad \Leftrightarrow \quad \quad
 \langle \, \vec{\phi}^{\,\dag} \cdot \vec{\phi}\,\rangle
 ={4 N \over g_0} ~~ .
 \ee
Therefore this phase corresponds to those field configurations
(\ref{eqn:GribovSet}) where the residual gauge invariance manifests
itself. In contrast, the Higgs phase of scalars corresponds to the field
configuration with vanishing vacuum expectation value of
$\vec{\phi}^{\,\dag} \vec{\phi}$ and thus the residual gauge invariance
does not play a role. Choosing the Lorentz gauge
\be
 S_{GF}={N\epsilon \over 12 m^2 g_0}\int_x\(\partial_{\mu}A^{\mu}\)^2
 ={N\epsilon \over 12 m^2 g_0}\int_p p_{\mu}p_{\nu}A^{\mu}(p)A^{\nu}(-p) \,.
 \ee
and combining it with (\ref{eqn:CPNquadr-action}), leads to the following propagator
 \be
 D_{\mu\nu}^{(2)}(p)={6 g_0 \over N\epsilon }
 \[ {m^2 \over p^2} \, {p_{\mu}p_{\nu} \over p^2} + {1 \over f^{(2)}}\(\delta_{\mu\nu}-{p_{\mu}p_{\nu} \over p^2}\) \]
 ~,
 \ee
where
 \be
 f^{(2)}\equiv\({p \over m}\)^{\epsilon} \({1\over 4}+{m^2\over p^{\,2} }\)^{\epsilon/2-1}
 \ _{2}F_1\({3\over 2} ~,~ {2-\epsilon\over 2} ~,~{5\over 2} ~;~ {1\over 1+4 \, m^2/p^2}\)
 ~.
 \ee
In the low-energy limit $p<<m$ one obtains
 \be
 D_{\mu\nu}^{(2)}(p) \underset{p<<m}\longrightarrow {6 m^2 g_0 \over N\epsilon }
 ~{\delta_{\mu\nu} \over p^2} ~.
 \ee

Using (\ref{eqn:prop-expansion}) the full $s-$channel of the
connected one-particle irreducible amputated four-point function is
then obtained by adding
 \be
 \Gamma_{s}^{(r)}\Big|_{A^2} = -(p_1-p_2)^{\mu} D_{\mu\nu}^{(r)}(s)(p_3-p_4)^{\nu}~,
 \ee
to (\ref{eqn:scalar-prop-vertex1}),(\ref{eqn:scalar-prop-vertex2}).
For the Higgs phase we get
\begin{equation}
 \Gamma_{s}^{(1)}\(p_i^2=0\)={-g_0 \over N(2+g_0 s I_0(s))}\bigl(p_1\cdot p_2+p_3\cdot p_4\bigr)
 + {g_0 \over 2N}{1\over 1+g_0f^{(1)}s^{\,\epsilon/2}}(p_1-p_2)\cdot(p_3-p_4)
 ~ .
 \label{eqn:massless-prop-vertex}
\end{equation}

We next adopt the approach of \cite{Bardeen:1976zh}, and show that
$g_c$ is an ultraviolet fixed point and the system is within the Higgs phase
if $g<g_c$ and within the confinement phase if $g>g_c$, where $g$ is the
renormalized running coupling constant defined by
 \be
 g_0=g Z_{g}\, \mu^{-\epsilon} ~,
 \ee
which renders $g$ dimensionless by introducing an arbitrary scale
$\mu$.

We impose the following renormalization condition
 \be
 \Gamma_{s}^{(1)}(\lambda;s=\mu^2,p_i^2=0)= - {g\,\mu^{-\epsilon}\over 2N}\bigl(p_1\cdot p_2+p_3\cdot p_4
 - (p_1-p_2)(p_3-p_4)\bigr)
 ~.
 \ee
Combining (\ref{eqn:massless-prop-vertex}) with the above condition
leads to
 \be
 {Z_g \over 1+ (g/g_c)Z_g}=1 \quad\Rightarrow\quad Z_g=\(1-{g \over g_c}\)^{-1}~.
 \label{eqn:Zg}
 \ee
Since the bare coupling $g_0$ is scale independent, we get
 \be
 \mu{d g_0 \over d\mu}=0\quad\Rightarrow\quad
 \beta_g=\mu{d g \over d\mu}=\epsilon g \(1-{g \over g_c}\)
 \quad .
 \label{eqn:beta-function}
 \ee

Moreover if $g<g_c$ it follows that $Z_g>0$ and thus one must choose
the Higgs phase as the solution to the gap equation
(\ref{eqn:CPNgap-eqn}), since the massive solution in
(\ref{eqn:CPNphases}) would be self-contradictory. On the other
hand, for $g>g_c$ the theory exists only in the confinement phase, since
if we let $g>g_c$ in (\ref{eqn:massless-prop-vertex}), there would
appear a tachyon pole at
 \be
 s=\mu^2\(1-{g_c \over g}\)^{2/\epsilon}
 \quad .
 \ee

Before turning to analyze the properties of the system at the fixed point, we wish to discuss a subtlety which is there in the $CP^{N-1}$ model both in $d=2$ and $d=2+\epsilon$ dimensions. This will accompany the analysis also in the other cases discussed later in this work. In the phase where the global symmetry is unbroken for $d=2+\epsilon$, which is the only phase at $d=2$, one notes that the IR limit and the large $N$ limit do not commute. This could leave one with the impression that one is free to make a choice of the way in which to order the limits. This is wrong. One clue comes from the $d=2$ case, taking first the large $N$ limit suggests that the theory describes 2N massive particles in the vector representation of $O(2N)$ if this was the case the global $SU(N)$ symmetry apparent in equation (\ref{eqn:coset-patt}) would not have been restored. This contradicts, in particular, the consequences of Coleman's theorem according to which one would expect that the lowest excitations of the system to be massive particles in the adjoint representation of $SU(N)$ or at least they should be neutral under $Z(N)$ \cite{Haber:1980uy}. Having the lowest mass particles in the fundamental representation would require them to be non perturbative states of such solitons which does seem likely. A Maxwell term could have allowed to bind the states back into the appropriate representations of $SU(N)$ but it is down by $1/N$. The correct description  of the physics should thus include Maxwell's term even though its presence is non leading in the $1/N$ expansion. The necessity to follow this order of limits can also be inferred from a more general point of view.

In general, once obtaining the leading term of the effective action in the $1/N$ expansion, one should inspect that it contains all the relevant operators allowed by the symmetries of the problem. Note, that anomalous dimensions encoded in the theory have to be taken into account when determining whether a given term is relevant or not. If allowed operators are absent, one should continue the expansion to the higher orders in $1/N$ until all the missing terms show up.  To ensure the stability of the theory, it is essential to add all such terms to the effective action even if the first time they appear is at a higher order in $1/N$ than the leading contribution\footnote{We thank David Kutasov for sharpening this understanding.}.

\subsection{The fate of the Maxwell term.}
\label{sec:CPn-Maxwell-gen}

In the current subsection we investigate the
generation and subsequent fate of the Maxwell term when the system
starts in the confinement phase and is gradually driven into a fixed
point.

The long distance behavior of (\ref{eqn:CPn-Aeff-action}) in the
confinement phase, $p<<m$, is given by
 \be
 \mathcal{O}(A^2) \simeq {\epsilon\,N \over 24 g_0 m^2}\int_x  F_{\mu\nu} F^{\mu\nu}~.
 \label{eqn:Maxwell-term}
 \ee
where (\ref{eqn:CPNphases}) has been used. As a result, the Maxwell term for the auxiliary
field $A_{\mu}$ is generated. In order to figure out what happens to
this term at a fixed point, note first that according to
(\ref{eqn:CPNphases}) and (\ref{eqn:Zg}) we get for the mass in the
strong-coupling phase
 \be
 m^2=\mu^2 \[  \(1-{g_c \over g}\) {\Gamma^{\,2}(1+\epsilon/2) \over \Gamma(1+\epsilon)}
 \]^{2/\epsilon} \quad.
 \label{eqn:mass-of-bosons}
 \ee

Hence, if we set $g=g_c$ ( with $\mu$ being fixed), we get $m^2=0$ and thus the condition $p<<m$ is not valid anymore. In particular, (\ref{eqn:Maxwell-term}) must be modified.
Relying on (\ref{eqn:CPn-Aeff-action}), we get in this case
\begin{multline}
 \mathcal{O}(A^2) \underset{m^2\rightarrow 0}\longrightarrow
 {N \over 2 \, (16\pi)^{\epsilon+1 \over 2}}
 {\Gamma[1+\epsilon/2] \Gamma[-\epsilon/2] \over \Gamma[3/2+\epsilon/2]}
 \int_p  A_{\mu}(p) A_{\nu}(-p)
 \[ \delta^{\mu\nu}p^{\,2} - p^{\,\mu}p^{\,\nu} \] \, p^{\,\epsilon-2}
 \\
 = {\sqrt{\pi}\over 16(2\pi)^{\,2+\epsilon}} ~ {\Gamma(\epsilon/2)\Gamma(\epsilon) \over \Gamma[\epsilon/2+3/2]}
 \int_x\int_y ~ {F^{a\mu\nu}(x) F_{\mu\nu}^a(y) \over |x-y|^{2\epsilon}} \quad .
 \label{eqn:CPn-critical-Max}
\end{multline}
Therefore neither the Maxwell term nor the corresponding
long-range force appear in this case.

Alternatively, one could start at some definite value of the
coupling constant $g(\mu_0)>g_c$ and gradually increase the energy
scale $\mu$. Solving (\ref{eqn:beta-function}) leads to
 \be
 \({\mu_0 \over \mu}\)^{\epsilon}\(1-{g_c \over g(\mu_0)}\)=1-{g_c \over g(\mu)}
 \quad .
 \ee
Substituting this result back into (\ref{eqn:mass-of-bosons}) yields a finite mass generated by the dimensional transmutation
 \be
 m^2=\mu_0^2 \[ \(1-{g_c \over g(\mu_0)}\) {\Gamma^{\,2}(1+\epsilon/2) \over \Gamma(1+\epsilon)}\]^{2/\epsilon} \quad.
 \ee

However, even though the mass is finite, when $\mu/\mu_0\rightarrow\infty$ the coupling constant is
driven to the critical value $g_c$ while the typical momentum
satisfies $p>>m$. This in turn leads to the same result \reef{eqn:CPn-critical-Max} and the Maxwell term does not emerge.

\section{The 3D gauged and ungauged $\phi^6$ $O(N)$ vector model.}
\label{sec:scalarCS}

In this section we investigate whether a Maxwell term is generated
in the case when the gauged system does possess a scale/conformal symmetry that is spontaneously broken. It turns out that the cases of spontaneous breaking of scale invariance and asymptotically free theories are very similar as far as their generating Maxwell terms for the gauge particles is concerned. We use the $\phi^6$ vector model in which the non-perturbative dynamics can be studied for a purely bosonic CFT  \cite{Bardeen:1983rv}\cite{Weinberg:1997rv}. Before proceeding to the various issues associated with the gauged system only, we turn to refine the properties which are shared by both the gauged and the ungauged cases. In particular, we reexamine the phase with spontaneous breaking of scale invariance where the massless dilaton emerges and a confining logarithmic potential is generated. The latter feature of confinement by the dilaton exchange was not brought up in previous investigations  \cite{Bardeen:1983rv}. We explore it in this work.

Recall that the large $N$ limit of the $O(N)$ $\phi^6$ is a very useful setup to study the exact behavior of conformal theories in higher than
two dimensions. It does have the limitation that the large $N$ limit does not commute with the limit of removing the
ultraviolet cutoff of the theory. To obtain a scale invariant theory of only scalars in this dimension one needs to take the
large $N$ limit first for a fixed UV cutoff and only then remove the cutoff. As shown in \cite{Bardeen:1983rv}, only in this particular order of limits, the theory is scale invariant and exhibits a phase with spontaneously broken scale invariance. A Goldstone massless particle, the dilaton, emerges. However, such a massless particle in 3D generates a long distance confining logarithmic potential which is accompanied by a $1/N$ overall coefficient, and thus also the IR limit does not commute with the $1/N$ expansion. Taking first $N$ to infinity may seem to lead to a free theory containing one massless particle and $N$ massive particles in the vector representation of $O(N)$. In the other order of limits the lowest massive excitations are the symmetric and singlet representations of $O(N)\times O(N)$.

This intricacy resembles the situation in the $CP^{N-1}$ model in $d=2$, where the IR and the large $N$ limits do not commute either\footnote{Another manifestation of this was found on the lattice the strong coupling and the large $N$ limit do not commute \cite{Rabinovici:1980dn}.}. However, there is a difference between the two: in the $CP^{N-1}$ model the large $N$ and the UV limits do commute, and thus the leading contribution to the force due to the gauge field can be reliably obtained although it comes with a $1/N$ overall coefficient.

Nevertheless, in light of the discussion at the end of subsection \ref{sec:CPn-critical-point} one may attempt to adopt the following procedure. First remove the IR cutoff and only next resort to the large $N$ limit and lastly remove the UV cutoff. Therefore in the phase where the scalars $\vec\phi$ are massive and the massless dilaton is formed, one can consider a non relativistic limit in which the potential picture makes sense, as it did in 2d $CP^{N-1} $ model, and the properties of the theory can be investigated as if the binding force operates in a CFT arena.

The dilaton on its own would form massive two-body bound states of the fields $\vec\phi$, some of them would be the lowest massive excitations of the system. As the bound states have no ``neutrality" properties one could expect also higher mass more complicated bound states. We are aware that this order of limits may result in less than logarithmically confining potential.

The uncovering of the ABJM models \cite{Aharony:2008ug} has led to a discovery of a large class of three dimensional scale invariant field theories. Some of these contain massless dilatons \cite{Mukhi:2008ux}. In those cases which are weakly coupled the dilaton couples to itself like a Goldstone boson and transmits a logarithmically confining potential among massive excitations\footnote{We thank Ofer Aharony for a discussion on this point.}. This without obviously suffering from the necessity to define a certain order of limits.

Next let us anticipate the results in the case when one gauges a $U(1)$ subgroup of the global symmetry without introducing a Maxwell term. This can be done for an $O(2N)$ global symmetry. The details are presented in the next sections. Although in $d<4$ the gauge coupling is dimensionful and the Maxwell term is relevant, in the absence of a Maxwell term the conformal symmetry is maintained.

The emergence of a Maxwell term presents the similar challenge to the one described above when studying the long distance forces. The value of the effective electric charge strictly vanishes for infinite $N$. Hence, like in the ungauged $\phi^6$ for infinite $N$ the degrees of freedom do not interact. For finite $N$ the theory is not conformal and the analysis is invalid.

However, once the Maxwell term emerges , the large $N$ and the IR limit need to be taken in the same order as in the case of the ungauged $\phi^6$. In the absence of the CS term the gauge field will confine at least logarithmically the U(1) charge and only the bound states which are neutral under $U(1)$ will be formed. They will be in the adjoint representation of $SU(N)$.

When the CS term is introduced either ab-initio or it emerges, it will be only the dilaton which provides a confining potential. The CS term on its own has no bulk degrees of freedom, but when coupled to a Maxwell term it effectively generates a mass for the gauge field. Hence, the photon becomes massive and the long-range force it carried disappears.

\begin{figure}[t!]
\centering \noindent
\includegraphics[width=8cm]{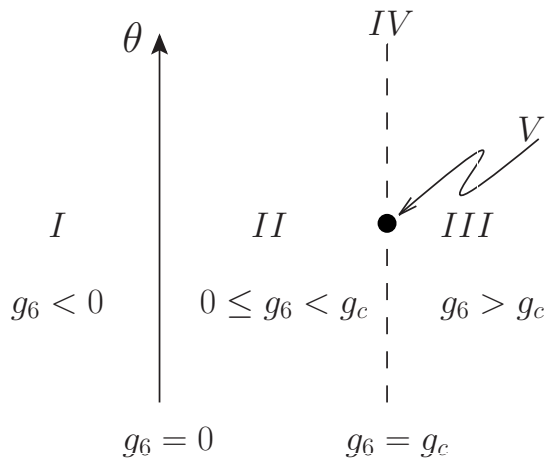}
\caption[]{The diagram of phases in the $\theta-g_6$ plane.
$g_c=(4\pi)^2$ represents the critical value of the coupling
constant. Phases $I$ and $III$ are unstable as argued in \cite{Bardeen:1983rv}.
Phase $II$ corresponds to a massless conformal phase, where the Maxwell term does not emerge. The critical line $g_6=g_c$ represents two distinct phases of the spontaneously broken scale invariance: the dashed line (phase $IV$) is associated with the phase where $\theta\neq 0$ and the particles are bound by dilaton only, whereas the bold point (phase $V$) corresponds to $\theta=0$ where only the states which are neutral under $U(1)$ will be formed.} \label{fig:Phase-diagram}
\end{figure}

As a result, the CS parameter splits the phase with spontaneously broken scale invariance into two distinct phases: the phase where only the dilaton binds the particles into an irreducible representation of $O(2N)\times O(2N)$ and the phase with neutral bound states belonging to an irreducible representation of $SU(N)$. Figure \ref{fig:Phase-diagram} demonstrates the enrichment of the phase structure when the CS term is considered. Phases $I$ and $III$ in the figure are unstable \cite{Bardeen:1983rv}, whereas phase $II$ corresponds to a massless conformal phase without a Maxwell term being generated.

There has been a suggestion \cite{Klebanov:2002ja} to relate a $d=3$ $O(N)$ CFT vector model on the boundary to higher spin bulk theories. It was pointed out \cite{Elitzur:2005kz} that in order to maintain, as needed, only the $O(N)$ singlet sector, one needs to study the IR limit of the $\phi^6$ vector model. At its critical point when the scale invariance is spontaneously broken the IR limit  consists of only one massless field - the $O(N)$ singlet dilaton. This remains also in the cases studied here. If one wants to have a theory containing only $O(N)$ singlets both massive and massless, one needs to gauge the full global symmetry so that the states in the theory are all singlets.

\subsection{The gap equations.}

The gauged 3-dimensional model considered in this section is given by
\begin{equation}
 S\left( \vec{\phi}\, , \, \vec{\phi}^{\,\dag} , A\right) =\int \[\overline{D_{\mu}\phi_i} \, D ^{\mu }\phi_i
 + {g_6 \over 3 N^2} (\phi_i\phi_i^{\dag})^3 + {N\theta \over 2} \, i \, \varepsilon^{\al\mu\nu}A_{\al}F_{\mu\nu}\]\, d^{3}x \, ,
 \label{eqn:gauged-scalar-action}
\end{equation}
The ungauged case and its supersymmetric extension were extensively
studied in a number of works \cite{Weinberg:1997rv}. In particular,
the phase with spontaneously broken scale invariance was explored in
\cite{Bardeen:1983rv,Bardeen:1984dx}. Recently, time-dependent
rolling of the system in the conformal potential unbounded from
below was solved exactly in the large $N$ limit \cite{Asnin:2009bs}.
Moreover, the effects of the CS coupling on the high-energy behavior
of the model was considered in \cite{Park:1994sw}. On the other
hand, the long wavelength physics of the model was not analyzed and
this is the main purpose of the current section. Without the CS term
the action is invariant under reflections. Therefore integrating out
various degrees of freedom will not introduce the CS term into the effective
action unless it appears in the model from the very beginning.

We demonstrate that once the system is in a phase with
spontaneously broken scale invariance, the Maxwell term for the
dummy gauge field $A_{\mu}$ is generated. While it is shown that the CS
term does not alter the saddle-point equations, it
does affect the long-distance physics. In particular, it screens that part of the confining potential for which the gauge field is responsible. It does so by introducing a mass for the gauge field. As an outcome, the bound states of the system with and without the CS term will not be the same. In fact, there will be less bound states in the presence of a CS term if at all since the long-range force associated with the Maxwell term does not confine in this case. However, without the CS term, only neutral states will be present in the spectrum.

The generating functional of this model is given by
\begin{equation}
 Z\left[ \vec{J}, \, \vec{J}^{\,\dag}\right] = \int D\vec{\phi} \,D\vec{\phi}^{\,\dag}\,DA_{\mu}
 \exp \left[ -S\left( \vec{\phi}\, , \, \vec{\phi}^{\,\dag} , A\right)
 - N S_{GF}(A)+\int d^{3}x \(\vec{J}^{\,\dag}\cdot \vec{\phi } + \vec{J}\cdot \vec{\phi}^{\,\dag}\) \right]  \, .
\end{equation}
where a gauge fixing term, $S_{GF}(A)$, was introduced into the
action in order to make the generating functional and consequently
the Green's functions well-defined. We choose the
Lorentz gauge condition
 \be
 S_{GF}(A)={1 \over 2\al} \int d^{3}x (\partial_{\mu}A^{\mu})^2 \, .
 \ee
Note also, that in our conventions the dimension of the arbitrary
parameter $\al$ is $1$. Hence, in what follows we choose the Landau
gauge, $\al\rightarrow 0$, in order to eliminate the unphysical
scale associated with $\al$.

Inserting (\ref{eqn:identity}) and integrating over $\phi_i \, , \,
\phi_i^{\,\dag}$ yields
\begin{equation}
 Z\left[  \vec{J}, \, \vec{J}^{\,\dag} \right]= \int D\rho D\lambda\,DA_{\mu}
 \, e^{-N S_{eff}(\rho,\,\lambda, \, A) }e^{ \int \vec{J}^{\,\dag}(x)\left(
 -D_{\mu}D^{\mu}+i\lambda\right)_{xy}^{-1}\vec{J}(y)d^{3}xd^{3}y }
 \, ,
 \label{eqn:partition}
\end{equation}
where
\begin{equation}
 S_{eff}(\rho,\lambda, A)=\int\[ {g_6 \over 3 }\rho^3 - i\lambda\rho
 + {i \theta \over 2} \, \varepsilon^{\al\mu\nu}A_{\al}F_{\mu\nu}+{1 \over 2\al} (\partial_{\mu}A^{\mu})^2\]d^{3}x
 + \texttt{Tr}\ln \left( -D_{\mu}D^{\mu} +i\lambda \right)  \, .
 \label{eqn:Seff}
\end{equation}

When $N$ is large, the saddle point method is used. In the Landau gauge for the Lorentz invariant vacuum
$\langle A_{\mu}\rangle=0$ there is no gauge field
contribution to the gap equations. Hence, varying the effective
action with respect to the auxiliary fields $\rho$ and $\lambda$,
yields\footnote{We use here the definition $\texttt{Tr}=\int
d^3x\,\texttt{tr}$}

\begin{equation}
 i\overline{\lambda} = g_6 \overline{\rho}^2 ~ ,\qquad
 \overline{\rho}=  \, \texttt{tr}\frac{1}{-\del^2+i\overline{\lambda}}
 =-\frac{\sqrt{i\overline{\lambda}}}{4\pi}\equiv-\frac{m}{4\pi} \, ,
 \label{scalar_gap}
\end{equation}
where barred quantities denote the solution of the gap equation and
$m$ will assume a role of a mass. Moreover, we have used the
dimensional regularization procedure in order to define the
divergent loop. Reinserting the gap equation for $\rho$ into the gap
equation for $\lambda$ leads to
 \be
 m^2=g_6 \, {m^2 \over (4\pi)^2} \, .
 \label{eqn:phi6gap}
 \ee
The theory is in a conformal invariant phase when $0\leq g_6< (4\pi)^2$ or when $g_6= (4\pi)^2$ and one chooses $m=0$. The theory is in a spontaneously broken scale invariant phase for $g_6= (4\pi)^2$ and dynamically generated nonzero mass $m$. In the cases when the coupling is either greater than $(4\pi)^2$ or less than $0$ the theory is unstable as argued in \cite{Bardeen:1983rv}.

\subsection{Dynamical generation of the Maxwell term.}

The aim of this subsection is to examine if in the phase with
spontaneously broken scale invariance there is a dynamical
generation of the Maxwell term for the $U(1)$ gauge field introduced
above.

Consider the energy, $E$, is much less than
the dynamically generated mass $m$
 \be
 E<<m \, .
 \ee
Then according to equation (\ref{eqn:low-energy-Aeff-action}), once
the $3-$dimensional effective action (\ref{eqn:Seff}) is expanded in
the vicinity of the saddle point $\overline{\rho} \,
,\,\overline{\lambda}\, , \, A_{\mu}=0$, we get
 \begin{multline}
 S_{eff}(\rho,\lambda, A)=\int\[  {g_6 \over 3 }\rho^3 + g_6 \overline{\rho}\rho^2- i\lambda\rho
 + {i\theta \over 2} \, \varepsilon^{\al\mu\nu}A_{\al}F_{\mu\nu}+{1 \over 2\al} (\partial_{\mu}A^{\mu})^2\]d^{3}x
 \\
 +\frac{1}{2} \int d^{\,3}x \int d^{\,3}y \, G^{\,2}(x,y)\lambda(x)\lambda(y) \,
 + {1 \over 96\pi m} \int d^3 x
 ~ F^{\mu\nu} F_{\mu\nu} + \, . \, . \, ,
 \end{multline}
where
 \be
  G(x-y)=\int {d^3p \over (2\pi)^3} {e^{ip(x-y)} \over p^2+m^2}
 \quad .
 \ee
and the constant terms associated with $\overline{\rho} \, ,\,\overline{\lambda}$ are omitted from \reef{eqn:CS-Aeff}, whereas
ellipsis there denote various interactions of the gauge field $A_{\mu}$
with the scalar field $\lambda$ and their self-interactions. To diagonalize the quadratic part of the effective action, let us apply the following shift
 \be
 \rho\rightarrow \rho+{i\lambda \over 2 g_6 \bar\rho}
 \, ,
 \ee
then
 \begin{multline}
 S_{eff}(\rho,\lambda, A)=\int\[  {g_6 \over 3 }\Big(\rho+{i\lambda \over 2 g_6 \bar\rho}\Big)^3 + g_6 \overline{\rho}\rho^2+ {\lambda^2 \over 4g_6\bar\rho}
 + {i\theta \over 2} \, \varepsilon^{\al\mu\nu}A_{\al}F_{\mu\nu}+{1 \over 2\al} (\partial_{\mu}A^{\mu})^2\]d^{3}x
 \\
 +\frac{1}{2} \int d^{\,3}x \int d^{\,3}y \, G^{\,2}(x,y)\lambda(x)\lambda(y) \,
 + {1 \over 96\pi m} \int d^3 x
 ~ F^{\mu\nu} F_{\mu\nu} + \, . \, . \, ,
 \label{eqn:CS-Aeff}
 \end{multline}
There are several effects associated with the
spontaneously broken scale invariant phase. A mass is generated for the scalar particles\footnote{The $\phi$-propagator is obtained by differentiating the partition function (\ref{eqn:partition}) with respect to the source $\vec J$ and therefore (to leading order in $1/N$) $i\bar\lambda=m^2$ plays the role of the physical mass.} and the Maxwell term is generated for the gauge field (\ref{eqn:CS-Aeff}).
The effective charge of the particles is fixed by the
dynamically generated scale, and according to (\ref{eqn:CS-Aeff}),
is given by $\sqrt{24\pi m/N}$.

If, on the other hand, $0\leq g_6<g_c$, then according to \reef{eqn:phi6gap} $m=0$ and the Maxwell term does not emerge\footnote{Outside $0\leq g_6\leq g_c$ range the system is unstable  \cite{Bardeen:1983rv}. }. This time expanding around the saddle point and taking the limit $m\rightarrow 0$ will lead to a nonlocal gauge invariant term \reef{eqn:low-energy-Aeff-action2} with $d=3$. The long-range potential is weaker in this case and does not lead to a confinement.

\subsection{Confinement.}

In general, a massless particle in three spacetime dimensions  generates a logarithmic confining potential. For a compact $U(1)$ gauge symmetry, the nonperturbative effects turn the logarithmic confining potential into a linear confining potential \cite{Polyakov:1976fu}. The $U(1)$ symmetry in this problem is a subgroup of the compact group $SU(2)$. Nevertheless, for our purposes it is enough that the potential is confining.

In the previous section, we showed that in the phase with spontaneously broken scale invariance a massless gauge particle emerges, and thus it binds the scalar degrees of freedom into neutral states. Furthermore, since the scale symmetry is spontaneously
broken there is an associated Goldstone particle - the massless dilaton. Hence, the dilaton on its own would confine the particles as well. In the case of ungaged $O(N)$ vector model \cite{Bardeen:1983rv} the latter observation was not addressed and therefore we find it instructive to shed light on the confining phenomenon in the current manuscript.

In the effective action \reef{eqn:CS-Aeff}, the dilaton is represented by the scalar field $i\lambda$, and one can readily verify that $i\lambda$ is massless by examining its propagator
 \begin{multline}
   D_{\lambda}(x-y)=\left\langle i\lambda (x)\, i\lambda (y) \right\rangle =
 \frac{8\pi m}{N}\int \frac{d^{3}p}{\left( 2\pi \right) ^{3}}
 ~\frac{e^{ip\,(x-y)}}{1-8\pi m \, B(p)}
 \\
 \underset{p<<m}{=} \frac{96\pi m^3}{N}\int \frac{d^{3}p}{\left( 2\pi \right) ^{3}}
 ~\frac{e^{ip\,(x-y)}}{p^2}\(1+ {3\over 20}{p^2\over m^2}+\mathcal{O}({p^4\over m^4})\)
 ~ ,
 \label{eqn:dilprop}
 \end{multline}
where
 \be
 B(q)=\int \frac{d^{\,3}p}{\left( 2\pi \right)^{3}}G(p)G(p+q)
 ={1 \over 4 \pi q} \arctan \left( {q \over 2m} \right)~ .
 \label{eqn:bubble}
 \ee

In the absence of CS term ($\theta=0$) the gauge particle is massless. Hence, in two space dimensions both, both the dilaton and the gauge particle, contribute to the logarithmic potential which we now turn to compute.

Differentiating (\ref{eqn:partition}) with respect to the source $J$ and
$J^{\dag}$ twice and then setting $J$ to zero leads to a path integral with two insertions of $\left( -D_{\mu}D^{\mu}+i\lambda\right)^{-1}$.  Expanding each such factor around the
solution of the gap equation and keeping only linear terms in small perturbations\footnote{Higher order terms will contribute to the $1/N$ correction, since propagators of $\lambda$ and $A_{\mu}$ carry $1/N$ factor.} leaves us with the following expression for the four-point function (in what follows we adopt the notation of Appendix \ref{appx:bfncl-det-expan} and there is no summation on the repeated indices $a$ and $b$)
 \begin{multline}
 \langle \phi_b(x_1) \phi_b^{\,\dag}(x_2)\phi_a(x_3)\phi_a^{\,\dag}(x_4) \rangle
 = \int_w\int_u G(x_1-w)G(w-x_2)\langle  i\lambda(w)i\lambda(u)\rangle G(x_3-u)G(u-x_4)
 \\
 - \( {\partial \over \partial x_1^{\mu}}-{\partial \over \partial x_2^{\mu}} \)
 \({\partial \over \partial x_3^{\nu}}-{\partial \over \partial x_4^{\nu}}\)
 \int_w\int_u G(x_1-w)G(w-x_2)\langle  A^{\mu}(w)A^{\nu}(u)\rangle G(x_3-u)G(u-x_4)~~.
 \label{eqn:4pointfunc}
 \end{multline}

The $A^{\mu}$ propagator can be read off the quadratic part of the effective action (\ref{eqn:CS-Aeff}) and we get in the low
energy limit (recall that we consider now $\theta=0$ case)
 \be
 \left\langle A^{\mu} (x)\, A^{\nu} (y) \right\rangle \simeq
 \frac{24\pi m}{N}\int_p
 ~e^{ip\,(x-y)}~\frac{\delta^{\mu\nu}}{p^2}
 ~ ,
 \ee
where $\alpha=24\pi m$ has been used in the gauge fixing term.

\begin{figure}[t!]
\centering \noindent
\includegraphics[width=12cm]{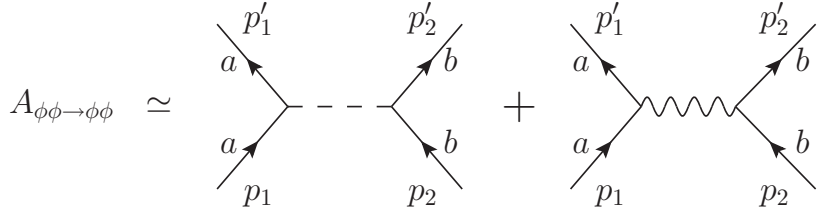}
\caption[]{Diagrams contributing to the nonrelativistic
$\phi\phi\rightarrow\phi\phi$ scattering amplitude. Exchange
diagrams are suppressed since the two interacting particles are
distinguishable. Dashed line represents propagator of the dilaton and thus stands for the auxiliary
field $\lambda$, whereas wavy line corresponds to the propagator of
the gauge field $A_{\mu}$.} \label{fig:diagrams}
\end{figure}

The diagrams in figure \ref{fig:diagrams} contribute to
the nonrelativistic $\phi\phi\rightarrow\phi\phi$ scattering
amplitude and thus according to (\ref{eqn:4pointfunc}) we get in this
limit
 \bea
 A_{\phi^+\phi^+\rightarrow\phi^+\phi^+}&=&A_{\phi^-\phi^-\rightarrow\phi^-\phi^-}
 \simeq \frac{96\pi m^3}{N}~\frac{1}{(p_1-p'_1)^2}
 +\frac{24\pi m}{N}~\frac{(p_1+p'_1)_{\mu}(p_2+p'_2)^{\mu}}{(p_1-p'_1)^2}
 = \mathcal{O}(m)\, ,
 \non
 A_{\phi^+\phi^-\rightarrow\phi^+\phi^-}
 &\simeq& \frac{192\pi m^3}{N}~\frac{1}{(\mathbf{p}_1-\mathbf{p}'_1)^2}~ .
 \eea
Hence, to leading order the electromagnetic repulsion of the similarly charged scalar particles is neutralized  by the attractive force due to the dilaton exchange. In contrast, if the particles are oppositely charged, then
they are confined by the logarithmic potential\footnote{The factor
$4m^2$ arises from the relativistic normalization conventions, and
must be dropped from the final result.}
 \be
 V(r)= {24 \over N} \,\, m \log (r) \, .
 \label{eqn:conf-potential}
 \ee
Hence, the bound states of the system must be neutral in this case.

However, it turns out that appearance of the $\theta$-term in the action \reef{eqn:CS-Aeff} changes the above conclusion. In the $\theta\neq 0$ case dealt within the next section, the gauge field is massive and does not confine in the IR. Thus, in this case the confinement is due only to the dilaton exchange. It is insensitive to the electric charge which is anyhow screened. The latter is reflected in a phase diagram \ref{fig:Phase-diagram}.

Before leaving this subsection it is instructive to examine the coupling of the dilaton to the original scalar degrees of freedom in \reef{eqn:gauged-scalar-action}. For this purpose, let us examine the leading $N$ effective action of the full theory in the low energy limit, see  the details in Appendix \ref{appx:eff-action}
 \begin{multline}
 \Gamma[\vec\phi, \psi, A_\mu]=\int_x \Big[ \vec{\phi}^{\,\dag}(-D_{\mu}D^{\mu}+m^2)\vec{\phi }
 +{1\over 2}\, \psi\Big(1+{3\over 20}{\del^2\over m^2}+\ldots\Big)(-\del^2\,\psi)
 +{1 \over 4}\,F^{\mu\nu} F_{\mu\nu}
 \\
 +\sqrt{{96\pi m^3\over N}} \,\vec{\phi }\cdot\vec{\phi}^{\,\dag}\psi
 + {i\theta \over 2} \, \varepsilon^{\al\mu\nu}A_{\al}F_{\mu\nu}+{1 \over 2\al} (\partial_{\mu}A^{\mu})^2 \Big]+\ldots
 \end{multline}
where we have rescaled the Maxwell field and the coupling constants according to $A_{\mu}\rightarrow\sqrt{24\pi m/N}A_{\mu}$,  $\al\rightarrow24\pi m\,\al$, $\theta\rightarrow \theta/(24\pi m)$; the scalar field $\psi=N^{1/2}(96\pi m^3 )^{-1/2}i\lambda\,$ carries the canonical mass dimension $1/2$,  and its low energy propagator possesses a canonical form;  $D_{\mu}=\del_{\mu}+i\sqrt{24\pi m/ N}A_{\mu}$ is the covariant derivative, whereas the ellipsis denote the $1/N$ corrections.

As seen from the above expression, the interaction term between the scalars of the model and the emerging dilaton is a relevant operator of dimension $3/2$ and therefore, as argued in section \ref{sec:CPn-critical-point}, is essential for the stability of the system although it is subleading in $1/N$. The same is true for the gauge coupling.

Apparently, the above expression is singular in the limit $m\rightarrow 0$. However, in this case one has to examine the full propagator \reef{eqn:dilprop} instead, since neither the limit $p<<m$ nor the rescaling of $i\lambda$ are legitimate. In fact, from \reef{eqn:dilprop} we learn that in this limit the dilaton (as expected) disappears.

\subsection{The mass of the gauge field in the presence of a CS term.}

Let us now illustrate that the CS term introduces a mass also for a gauge
field whose Maxwell term was dynamically generated as it does when the Maxwell term is there ab-initio. This results in the screening of the logarithmic Coulomb potential computed in the previous section.

Building on (\ref{eqn:quad-Aeff-action}) and (\ref{eqn:Seff}) we get
the following expression for the full quadratic in $A_{\mu}$ piece
of the effective action
\begin{multline}
 \mathcal{O}(A^2)= N \int_p   A_{\mu}(-p) A_{\nu}(p)
 \Big[~ {p^2\delta^{\mu\nu}-p^{\mu}p^{\nu} \over 16\pi p}\(  -2{m \over p} + {p^2+4m^2 \over p^2}\arctan {p \over 2m} \)
 \\
 +{1 \over 2\al} p^{\mu}p^{\nu}-\theta \, \varepsilon^{\mu\al\nu}p_{\al}~\Big] \, .
\end{multline}
The propagator of the photon is given by the inverse of the
quadratic part
 \be
 D_{\mu\nu}(p)={1\over N}\,{\Gamma \over \Gamma^2+8\,\theta^2p^2} \[ \delta_{\mu\nu}-{p_{\mu}p_{\nu}\over p^2}
 - 2\,\theta \, {\varepsilon_{\mu\nu\al} \, p^{\al} \over \Gamma}
 -{\al \over \Gamma}\(  8\theta^2 - {\Gamma^2 \over p^2 }\){p_{\mu}p_{\nu}\over p^2}\] \, ,
 \label{eqn:photonprop}
 \ee
where
 \be
 \Gamma(p^2)=-{m \over 4\pi} + {p^2+4m^2 \over 8\pi p}\arctan {p \over 2m} ~ .
 \ee
Expanding this propagator in the low energy limit $p<<m$ and
applying the Landau gauge $\al\rightarrow 0$, yields
 \be
 D_{\mu\nu}(p)\simeq{24\pi m \over N}{1 \over p^2 + M_{\gamma}^2} \[ \delta_{\mu\nu}-{p_{\mu}p_{\nu}\over p^2}
 - \,\, {M_{\gamma} \over \sqrt{2}} \,\, {\varepsilon_{\mu\nu\al} \, p^{\al} \over p^2} \] \, ,
 \label{eqn:massive-Aprop}
 \ee
where $M_{\gamma}=48\sqrt{2} \, \pi m \, \theta$.

Due to the presence of the CS term the gauge field becomes massive with
$M_{\gamma}$ being its mass. As a result, the CS term screens the Coulomb
part of the spatial confining potential and we get
 \be
 V(r)={12 \over N} \,\, m \log (r)\pm {12 m\over N} \,\, K_0 (r M_{\gamma}) \underset{r
 M_{\gamma}>>1}{\longrightarrow}
 {12 \over N} \,\, m \log (r) \pm {m \over M_{\gamma}} ~ {e^{-r M_{\gamma}} \over r  }\, ,
 \ee
where upper (lower) sign corresponds to the similarly (oppositely) charged particles.
Note that when $\theta\rightarrow 0$ we recover the previous result,
since in the vicinity of zero we have $K_0 (x)\simeq - \log x$.

\section{A supersymmetric extension of the 3D gauged $\phi^6$ model.}
\label{sec:SUSY-CS}

The main goal of the current section is to demonstrate that most of the
properties studied in the previous sections are maintained in the
presence of SUSY and in addition some new features arise. In
particular, using the example of the massive $\mathcal{N}=1$ gauged 3D SUSY model we show that
the generation of the Maxwell term is not prevented by supersymmetry.
Moreover, the CS term emerges dynamically as well, and
there is no need to introduce it by hand into the action. The latter is
a consequence of the fact that a fermion mass term is not
invariant under reflection and violates parity conservation in odd dimensions.
As a result, the long-range $U(1)$ force is screened and the confinement is entirely due to the dilaton and its superpartner dilatino. The Maxwell and the CS terms are accompanied by the superpartners that are generated
dynamically: the gaugino kinetic and mass terms respectively.

Furthermore, in the case of the supersymmetric extension of the $\phi^6$
model studied in the previous section, we illustrate that all these terms are dynamically generated if the system is
in a special phase where the scale invariance is spontaneously
broken \cite{Bardeen:1984dx}. In particular, the parity violation emerges
dynamically in this case as it is driven by dynamically generated fermion mass.

\subsection{$\mathcal{N}=1$ 3D gauged massive SUSY.}

First we show that for a massive $\mathcal{N}=1$ gauged SUSY theory the Maxwell term emerges. Consider a complex scalar superfield
 \be
 \Phi(x,\theta)=A(x)+\theta\psi(x)-\theta^2F(x)~,
 \ee
with $A,\,\theta^\al$ and $\psi^\al$ being respectively complex
scalar, real Majorana two-component spinor and a two-component
complex spinor, while $F$ is a complex auxiliary scalar
field\footnote{The notations and conventions adopted throughout this
section are those of \cite{Gates:1983nr}. The representation for the
$\gamma$-matrices is taken as
$\gamma^{\mu}=(\sigma_y,i\sigma_z,i\sigma_x)$ and
$\{\gamma^\mu,\gamma^\nu\}=-2\,\eta^{\mu\nu}$, where
$\gamma^{\mu}=(\gamma^{\mu})_\al^{~\,\beta}$, with the metric being
given by $\eta^{\mu\nu}=\mathrm{diag}(-++)$,
$\theta^2=1/2\,(\theta^\al\theta_\al)$ and for an arbitrary
two-component complex spinor $\bar\psi=(\psi^\al)^*$. First half of
the Greek letters $\alpha, \beta,...$ denotes spinor indices,
whereas second half $\mu,\nu,...$ stands for the Euclidean space
indices.}. We now examine the Lagrangian describing the minimal supersymmetric
coupling of a gauge vector particle to a complex massive scalar
multiplet
 \be
 S=-{1 \over 2}\int d^3x\,d^2\theta \[ (D^\al+i\Gamma^\al)\bar\Phi(D_\al-i\Gamma_\al)\Phi
 - 2\,m \bar\Phi\Phi \]
 ~ ,
 \ee
where $D_{\al}=\del_{\al}+i\theta^\beta\,\del_{\al\beta}$ is the covariant derivative on a superspace, and the real spinor gauge superfield is given by
 \be
 \Gamma^\al (x,\theta)=\chi^\al(x)-\theta^\gamma\[\delta_\gamma^{~\al} B(x)-iV_\gamma^{~\al}(x)\]
 -\theta^2\[2\lambda^\al(x)+i\partial_\gamma^{~\al} \chi^\gamma(x)\]
 ~ ,
 \ee
with $\chi^\al$ and $\lambda^\al$ being Majorana spinors, $B$ is a
real scalar, whereas
$V_\beta^{~\al}=V_{\mu}(\gamma^\mu)_\beta^{~\al}$ is a traceless
second-rank spinor corresponding to the vector gauge field $V_\mu$.

This action can be rewritten in terms of covariant components of
$\Phi(x,\theta)$ defined by covariant projection \cite{Gates:1983nr}
 \bea
 A'(x)&=&\Phi(x,\theta)|=A(x)~,
 \non
 \psi'_\al(x)&=&(D_\al-i\Gamma_\al)\Phi(x,\theta)|=\psi_\al (x)-i\chi_\al (x) A(x)~,
 \non
 F'(x)&=&(D_\al-i\Gamma_\al)^2\Phi(x,\theta)|=F(x)-iB(x)A(x)-i\chi(x)\psi(x)-\chi^2(x)A(x)
 ~ ,
 \eea
where vertical bar means evaluation at $\theta=0$. Omitting primes
for simplicity of notation, eliminating auxiliary field $F$ by using
its algebraic equation of motion and performing Euclidean
continuation, yields
 \be
 S=\int d^3x \[ -\bar\psi(\displaystyle{\not}\partial-i\displaystyle{\not}V+m)\psi-(i\bar\psi\lambda\,A+h.c.)
 -\bar A(\partial_\mu-iV_\mu)^2 A + m^2\bar A A  \]
 ~ ,
 \ee
where the Euclidean $\gamma$-matrices are taken as
$\gamma^{\mu}_E=-(\sigma_y,\sigma_z,\sigma_x)$ and
$\{\gamma_E^\mu,\gamma_E^\nu\}=2\,\delta^{\mu\nu}$.

We now assume that typical momentum is much less than the mass of
the complex scalar superfield and integrate out $\psi$ and $A$
fields to get an effective (Euclidean) theory of the gauge vector
particle and gaugino
 \be
 S_{eff}(V_{\mu},\lambda)=\texttt{Tr}\log\[-(\partial_\mu-iV_\mu)^2+m^2
 +\lambda(\displaystyle{\not}\partial-i\displaystyle{\not}V+m)^{-1}\lambda\]
 -\texttt{Tr}\log(\displaystyle{\not}\partial-i\displaystyle{\not}V+m)
 \,.
 \ee

\begin{figure}[t!]
\centering \noindent
\includegraphics[width=5cm]{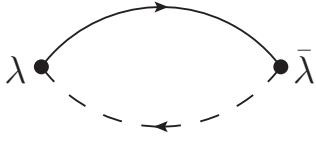}
\caption[]{Diagramatic representation of the second term in
(\ref{eqn:SUSY-eff-action}). Dashed arc corresponds to the
propagator of the complex scalar field $A$, solid arc stands for the
propagator of the complex fermion $\psi$, whereas bold dots are
associated with gaugino field $\lambda$ - a real two-component
Majorana fermion.} \label{fig:gaugino-term}
\end{figure}

Expanding this expression in the weak field approximation leads to
the following expression for the quadratic part of the effective
action
 \begin{multline}
 S_{eff}(\lambda,V_\mu)=\int {d^{3} p \over (2\pi)^3 }  V^{\mu}(-p)
 V^{\nu}(p) \,
 {p^2\delta_{\mu\nu}-p_{\mu}p_{\nu} \over 16\pi |p|}\(  -2{|m|\over |p|} + \(1+4{m^2\over p^2} \)\arctan {|p| \over 2|m|} \)
 \\
 +\int_x\int_y G(x-y)\bar\lambda(y)
 \langle y|(\displaystyle{\not}\partial+m)^{-1}|x\rangle\lambda(x)
 \\
 +\int_p V^{\mu}(-p)V^{\nu}(p)\,{p^2\delta_{\mu\nu}-p_\mu p_\nu \over 16\pi |p|}
 \[ 2{|m|\over |p|}+\(1-4{m^2\over p^2} \) \arctan {|p| \over 2|m|}\]
 \\
 -{m\over 4\pi}\,\varepsilon^{\al\mu\nu}\int_p V_{\mu}(-p)V_{\nu}(p)\,{p_\al\over |p|}\,\arctan {|p| \over 2|m|}
 +\ldots
 \label{eqn:SUSY-eff-action}
 \end{multline}
where the first term represents a contribution of the 3D bosonic
functional determinant (\ref{eqn:quad-Aeff-action}), the second term is associated with figure \ref{fig:gaugino-term}, whereas the third and
fourth terms emerge entirely from the fermionic functional
determinant (see Appendix \ref{appx:ffncl-det-expan} for details).

Using Feynman parametrization and dimensional regularization
(\ref{Appxeqn:dimreg-1}), (\ref{Appxeqn:dimreg-2}), yields
\begin{multline}
 \int_x\int_y G(x-y)\lambda(y)
 \langle y|(\displaystyle{\not}\partial+m)^{-1}|x\rangle\lambda(x)
 \\
 =\int_p \lambda(p)\int_k {m-i\displaystyle{\not}k \over k^2+m^2}
 {1 \over (p+k)^2+m^2}\lambda(-p)
 \\
 = -{1\over 8\pi}\int_p \lambda(p)\( i{\displaystyle{\not}p \over p}+2{m \over p} \)
 \lambda(-p) \arctan {|p| \over 2|m|}
  ~ .
\end{multline}
Combining altogether, we finally obtain
\begin{multline}
 S_{eff}(\lambda,V_\mu)=\int_p  V^{\mu}(-p)
 V^{\nu}(p) \, {p^2\delta_{\mu\nu}-p_{\mu}p_{\nu} \over 8\pi |p|}\arctan {|p| \over 2|m|}
 \\
 -{1\over 8\pi}\int_p \lambda(p)\( i{\displaystyle{\not}p \over |p|}+2{m \over |p|} \)
 \lambda(-p) \arctan {|p| \over 2|m|}
 \\
 -{m\over 4\pi}\,\varepsilon^{\al\mu\nu}\int_p V_{\mu}(-p)V_{\nu}(p)\,{p_\al\over |p|}\,\arctan {|p| \over 2|m|}
 +\ldots
 \label{eqn:SUSY-quad-eff-action}
\end{multline}

In the long wavelength limit $p<<m$, one can expand the integrands
in the above expression and get
 \be
 S_{eff}(\lambda,V_\mu)={1 \over 16\pi |m|} \int_x
 \[ {1 \over 2}F_{\mu\nu}F^{\mu\nu}-i\,m\,\varepsilon^{\al\mu\nu} V_\al F_{\mu\nu}
 -\lambda(x)(\displaystyle{\not}\partial+2m)\lambda(x) \]
 +\ldots
 \ee
To make SUSY manifest, we rewrite this action in terms of superfield
$\Gamma^\al$ as follows
 \be
 S_{eff}(\Gamma^\al)={1 \over 32\pi |m|}\int d^3x\,d^2\theta\( {1\over 4}D^\gamma D^\al\Gamma_\gamma+m\,\Gamma^\al \)
 D^\beta D_\al\Gamma_\beta
 +\ldots
 \ee

\subsection{Gauged $\mathcal{N}=1$ supersymmetric model in the large $N$ limit.}

Let us now explore the supersymmetric realization of the $SU(N)$
invariant $\phi^6$ model studied in the previous section and examine the emergence of the Maxwell term. In terms
of superfields the action of such a system is given by
 \be
 S={1 \over 2}\int d^3x\,d^2\theta \[-(D^\al+i\Gamma^\al)\bar\Phi(D_\al-i\Gamma_\al)\Phi
 + {g\over  N}(\bar\Phi\Phi)^2\]
 ~ ,
 \ee
here $\Phi$ is an $N$-component vector superfield and $\Gamma_\al$
is $SU(N)$ singlet.

In component form Euclidean counterpart of this action is given by
 \begin{multline}
 S=\int d^3x \bigl[ -\bar\psi(\displaystyle{\not}\partial-i\displaystyle{\not}V)\psi-(i\bar\psi\lambda\,A+h.c.)
 -\bar A(\partial_\mu-iV_\mu)^2 A
 \\
 +{g^2\over N^2}(\bar A\,A)^3 - {g\over  N} (\bar A\,A)(\bar\psi\psi)
 -{g\over 2\, N}(A\bar\psi+\psi \bar A)(A\bar\psi+\psi \bar A)\bigr]
 ~ ,
 \end{multline}
Using the following two identities
 \bea
 1\sim\int D\rho \, \delta(\bar A A-N\rho)
 &\sim& \int D\rho \, D\sigma \, e^{-i\int \sigma( \bar A A-N\rho )d^{3}x}
 \\
 e^{-\,{1\over 2}\,\xi M \xi}&=& (\det M)^{1\over 2}\int d\eta e^{-{1\over 2}\,\eta M^{-1} \eta + \eta\xi}
 ~ ,
 \eea
where $\xi,\eta$ are two auxiliary Majorana fields, yields
\begin{multline}
 S=\int d^3x \bigl[
 -\bar\psi(\displaystyle{\not}\partial-i\displaystyle{\not}V+g\rho)\psi+[\bar\psi(\eta-i\lambda)\,A+h.c.]
 -\bar A\[(\partial_\mu-iV_\mu)^2-i\sigma \]A
 \\
  +N\,g^2\rho^3-iN\rho\sigma-{N \over g}\,\eta^2\bigr]
 ~ ,
 \end{multline}
Integrating out $\psi$ and $A$, we obtain
\begin{multline}
 S_{eff}=N\texttt{Tr}\log\[-(\partial_\mu-iV_\mu)^2+i\sigma
 +(\eta+i\lambda)(\displaystyle{\not}\partial-i\displaystyle{\not}V+m)^{-1}(\eta-i\lambda)\]
 \\
 -N\texttt{Tr}\log(\displaystyle{\not}\partial-i\displaystyle{\not}V+g\rho)
 +N \int d^3x \bigl[\,g^2\rho^3-{1 \over g}\,\eta^2-i\rho\sigma \bigl]
 \,.
 \label{eqn:NSUSY-eff-action}
 \end{multline}

The last form of the action suggests a saddle point evaluation. The
Lorentz invariant gap equations are
 \bea
 \lambda&=&V_\mu=\xi=\eta=0~,
 \non
 i\bar\sigma&=&3g^2\bar\rho^2-g\,\texttt{tr}{1 \over \displaystyle{\not}\partial+g\,\bar\rho}~,
 \non
 \bar\rho&=&\texttt{tr}{1 \over -\partial^2+i\bar\sigma}
 ~ .
 \eea
These set of equations is unaltered by gauging, namely it is
identical to the ungauged case \cite{Bardeen:1984dx}. In particular,
the masses of the complex scalar and complex fermion are given
respectively by $m_A^2=i\bar\sigma$ and $m_\psi=g\,\bar\rho$,
therefore one can show that SUSY is maintained, i.e.
$m_A^2=m_\psi^2$. Furthermore, the theory is conformal and possesses
two $SU(N)$ invariant phases, one with $g\neq\pm(4\pi)$ or $g=\pm(4\pi)$ and
vanishing mass, and the other with spontaneously broken scale
invariance and a dynamically generated arbitrary mass
for\footnote{As mentioned, the theory violates parity. However, the space reflection is
equivalent to the change $g\rightarrow-g$, and this in turn reveals the origin
of the +/- sign above, see e.g. \cite{Bardeen:1984dx}.} $g_6=\pm(4\pi)$. In the latter phase the parity violation is dynamically generated.

Expanding (\ref{eqn:NSUSY-eff-action}) around the solution of
the gap equations, $\rho\rightarrow\bar\rho+\rho$,
$\sigma\rightarrow\bar\sigma+\sigma$, keeping only quadratic terms
in small perturbations and using
(\ref{eqn:bubble}),(\ref{eqn:SUSY-quad-eff-action}) one obtains
\begin{multline}
 S_{eff}=N\int_p  V^{\mu}(-p)
 V^{\nu}(p) \, {p^2\delta_{\mu\nu}-p_{\mu}p_{\nu} \over 8\pi |p|}\arctan {|p| \over 2|m|}
 \\
 -{N\over 8\pi}\int_p [\eta(p)+i\lambda(p)]\( i{\displaystyle{\not}p \over |p|}+2{m \over |p|} \)
 [\eta(-p)-i\lambda(-p)]\arctan {|p| \over 2|m|}
 \\
  +{g^2 N\over 8\pi}\int_p \rho(p)\rho(-p)|p| \(  2{|m|\over |p|} + \(1+4{m^2\over p^2} \)\arctan {|p| \over 2|m|} \)
 \\
 -{m\,N\over 4\pi}\,\varepsilon^{\al\mu\nu}\int_p V_{\mu}(-p)V_{\nu}(p)\,{p_\al\over |p|}\,\arctan {|p| \over 2|m|}
 \\
 +{N \over 2} \int_p {\sigma(p)\sigma(-p) \over 4 \pi |p|} \arctan \left( {|p| \over 2|m|} \right)
 \\
 +N \int d^3x \bigl[\,3g^2\bar\rho\rho^2-{1 \over g}\,\eta^2-i\rho\sigma \bigr]
 +\ldots
\end{multline}
here $m=m_\psi=m_A$.

Considering the phase with spontaneously broken scale invariance,
expanding the above result in the long wavelength limit $|p|<<|m|$
and rescaling the fields
 \bea
 \rho&\rightarrow&\tilde\rho\,\sqrt{3|m| \over 2 \pi}~,
 \non
 i\sigma&\rightarrow& 8\,|m|\, \tilde\sigma\sqrt{3 \pi |m|}
 ~ ,
 \eea
yields
\begin{multline}
 S_{eff}(\lambda,V_\mu)={N \over 16\pi |m|} \int_x
 \[ {1 \over 2}F_{\mu\nu}F^{\mu\nu}-i\,m\,\varepsilon^{\al\mu\nu} V_\al F_{\mu\nu}
 -\lambda(\displaystyle{\not}\partial+2m)\lambda-\eta\displaystyle{\not}\partial\eta \]
 \\
 +N \int d^3x \[-\tilde\rho\(\partial^2+6 m^2\)\tilde\rho
 -\tilde\sigma\( \partial^2+12 m^2 \)\tilde\sigma-12\sqrt{2}\,m^2\tilde\rho\tilde\sigma \]
 +\ldots
\end{multline}
The mass matrix of the second line possesses zero
eigenvalue associated with the Goldstone boson - the massless dilaton.
It necessarily appears in the spectrum since the model exhibits
spontaneous breakdown of the scale invariance. The last term in the
first line corresponds to the massless dilatino - the superpartner of
the dilaton. Finally, the first three terms represent dynamically
generated gauge vector particle, gaugino and their appropriate masses.

Note also, that since the gauge field is massive it does not confine in the IR, and thus the bound states emerge due to the logarithmic confining potential generated by either the dilaton or the dilatino. The former binds the particles having the same statistics and thus creates the bosonic bound states, whereas the latter binds the particles possessing different statistics and therefore is responsible for the generation of the fermionic bound states. This behavior is a reflection of SUSY.

\section {An anomalous 2D coset model.}
\label{sec:anom2D}

Requiring theories to be free of anomalies of local symmetries imposes constraints on the matter content of both gauge theories, general coordinate invariant ones and theories with local scale invariance.  One may be tempted to investigate the properties of such theories when they are left to be
anomalous. There are general ideas on what should go wrong, but this was not done yet explicitly for theories with Weyl or general coordinate invariance anomalies, it was done for some cases of gauge theories.

A theory with an anomalous gauge symmetry will behave differently in different gauges, this does not imply that
the theory is inconsistent in all gauge choices. Consistent refers to unitarity. In fact, in the gauge $A_0=0$ the theory is unitary, however it is expected that for anomalous theories  Lorentz invariance will not be restored as it is in the anomaly free theories. The idea that such a theory will be consistent is implied already in Dirac's work (his Yeshiva lectures \cite{Dirac:1964}). He discusses the possibility to implement in quantum mechanics a condition that both the coordinate operator $x$ and its conjugate momentum $p_x$ annihilate a state although their commutator is a c number. His answer is that indeed it may happen in the case when the Hamiltonian  has neither $x$ nor $p_x$ dependence. For an anomalous gauge theory what happens is that the commutator of Gauss's law at different points contains a c number.

Two dimensional anomalous gauge theories have been studied in some detail in \cite{Halliday:1985tg}, and we refer the reader to the original paper for many of the details. The anomalous Schwinger model is parameterized by an effective fermion carrying a right handed electric charge $e_R$ and a left-handed electric charge $e_L$. The model can be diagonalized in the gauge $A_0=0$ and its spectrum was obtained. It is indeed unitary and contains particles whose spectrum is not Lorentz invariant. These are described below.

Next let us point out that the Lagrangian realizations of $G/H$ coset models as mentioned above are conformal theories. Essentially the dynamics is that all states which are not massless have infinite mass in the absence of a Maxwell term for the gauge field, or alternatively in the presence of a Maxwell term the theory becomes very strongly coupled and flows in the infrared to the conformal $G/H$ theory. Care was always taken that the gauged group $H$ be anomaly free. This was also important for the geometrical interpretation of such systems. Here we ignore this warning and consider what would go wrong if one gauges an anomalous group $H$ in the coset construction. That would be tantamount to taking the strong coupling limit in the dispersion relations.

In the $A_0=0 $ gauge when $G$ and $H$ are both  $U(1)$ , what happens is that the states with a non relativistic dispersion relation obtain an
infinite energy and all what survives at finite energy are states with relativistic dispersion relations. The speed of light is  renormalized.
This we show below. Gauging an anomalous gauge passed with impunity. This may have interesting implications on the geometrical interpretations of the $G/H$ models leading also to the singular ones.

As shown in \cite{Halliday:1985tg}, the dispersion relation is determined by the positive solutions of the following equation
 \be
 E^2(E^2-k^2-e_L^2-e_R^2)^2-k^2(e_L^2-e_R^2)^2=0
 ~,
 \label{eqn:dispersion-rel}
 \ee
where $e_L$ and $e_R$ denote the total left handed and the total right handed electric charges in the underlying Schwinger model. For $e_L=e_R$ the system is anomaly free. The general solution of this equation is given by
 \be
 E^2={a^2 \over 18 u}-{a \over 3}+{u \over 2}~,
 \label{eqn:dispersion-solution}
 \ee
where
 \bea
 a&=&-2(k^2+e_L^2+e_R^2)~ ,
 \non
 u&=& \[4 k^2(e_L^2-e_R^2)^2+{a^3 \over 27}+k(e_L^2-e_R^2) \sqrt{16 \, k^2 \, (e_L^2-e_R^2)^2+{8 \over 27}\,a^3}\]^{1/3}
 ~ .
 \eea
The above expression for $u$ generates up to three distinct
solutions\footnote{There are three
cubic roots related by a factor which is one of the two non-real
cubic roots of one, and two square roots of any sign; but these 6
expressions can generate only 3 distinct solutions.} of (\ref{eqn:dispersion-rel}). In the
particular case $e_R=0$, three solutions are given by
 \bea
 E_{\pm}(k)&=&  \sqrt{{k^2 \over 4}+e_L^2}~\pm~ {k \over 2}
 ~ , \\
 E_{R}(k)&=& k
 ~ .
 \eea
In the limit $e_L>>k$ with $e_R = \al\,e_L + \gamma$, where $\al$ and $\gamma$ are fixed, the solutions
(\ref{eqn:dispersion-solution}) of (\ref{eqn:dispersion-rel}) can be expanded as follows
 \bea
 E_{\pm}&=&e_L\sqrt{1+\al^2} \pm {k \over 2}\,\Big|{\al^2-1 \over \al^2+1 } \Big|+ {\al \over \sqrt{\al^2+1}}\,\gamma +\mathcal{O}(k^2/e_L)~,
 \non
 E_{R}&=&k  \Big|{\al^2-1 \over \al^2+1 } \Big| +\mathcal{O}(k^2/e_L)
 ~ .
 \label{eqn:asympt-sol}
 \eea
Let us show how the original symmetry $e_L\leftrightarrow e_R$ of (\ref{eqn:dispersion-rel}) manifests itself in the above expression. Inverting the initial ansatz leads to
 \be
 e_L={1 \over \alpha}e_R-{\gamma\over\alpha}~,
 \label{eqn:eL-vs-eR}
 \ee
Substituting it into (\ref{eqn:asympt-sol}) yields
 \bea
 E_{\pm}&=&e_R\sqrt{1+\al^{-2}} \pm {k \over 2}\,\Big|{\al^{-2}-1 \over \al^{-2}+1 } \Big|- {\gamma \over \alpha^2\sqrt{\al^{-2}+1}} +\mathcal{O}(k^2/e_L)~,
 \non
 E_{R}&=&k  \Big|{\al^{-2}-1 \over \al^{-2}+1 } \Big| +\mathcal{O}(k^2/e_L)
 ~ .
 \eea
As expected from the symmetry $e_L\leftrightarrow e_R$, these solutions are obtained from (\ref{eqn:asympt-sol}) by applying the following  replacements (see (\ref{eqn:eL-vs-eR}))
 \be
 e_L\rightarrow e_R~,\quad \alpha\rightarrow {1\over\alpha}~,\quad \gamma\rightarrow -{\gamma\over\alpha}~.
 \ee

The same results can be obtained by expanding the cubic equation
(\ref{eqn:dispersion-rel}) rather than its full solution
(\ref{eqn:dispersion-solution}). Indeed, assuming $e_L>>k$ with $e_R
/ e_L = \al$ fixed ($\gamma=0$), yields
 \be
 E^2(k^2/e_L^2+\al^2+1-E^2/ e_L^2)^2-k^2(\al^2-1)^2=0
 ~ ,
 \ee
therefore to leading order in $k/ e_L$
 \be
 E^2=k^2 \({\al^2-1 \over \al^2+1 }\)^2 \quad\quad \mathrm{or} \quad\quad
 E^2=e_L^2(\al^2+1)
 ~ .
 \ee
Hence, in this particular limit the spectrum consists of a single massless particle in a relativistic theory with a modified speed of light. In the example above, the Maxwell term did not emerge just as it did not emerge when the guaging involved in a non anomalous system.

In contrast, in the limit $e_L>> k$ with $e_R - e_L = \gamma$
fixed  ($\al=1$), one gets
  \be
 E^2\({E^2 - k^2 \over 2 e_L + \gamma}-{\gamma^2 \over 2 e_L + \gamma}-{e_L+\gamma \over 1 + \gamma/ 2 e_L} \)^2-k^2\gamma^2=0
 ~,
 \ee
therefore to leading order in $k/ e_L$, the solution is either zero
or $E^2 \simeq \, 2 e_L^2$, and thus the model exhibits no finite
non trivial spectrum in this particular limit.

\section{Concluding Remarks.}

We have studied a variety  of gauge invariant theories without a Maxwell term.  In those theories in which the gauge coupling carried dimensions we found that the term was generated essentially when expected. That is whenever the theory had a scale be it generated dynamically, be it formed in an asymptotic free theory  or be it present ab-initio.

Of particular interest to us was the case when the scale symmetry was broken spontaneously and the low energy  spectrum consisted only of a massless singlet field of the appropriate group - the dilaton. $\mathcal{N}=1$ supersymmetry did not obstruct this feature. Whenever a scale was absent the Maxwell term failed to emerge. We have studied in this context aspects of the structure of gauge theories with and without the initial presence of a Chern-Simons term. Some interesting patterns emerged.

In four dimensions the coupling is classically dimensionless and the absent Maxwell term is a classically marginal operator. One way to view its absence which is useful  in lower dimensions is to consider this case as the infinite gauge coupling limit. In lower dimensions this is valid as even in the ab-initio presence of a Maxwell term the coupling becomes infinite in the deep IR. In the IR non trivial conformal theories, we saw that as long as the scale symmetry is not broken the Maxwell term does not emerge, while in an asymptotically free theories the coupling is not a free parameter and the term emerges.

Another way to consider this is in the strong coupling limit of a lattice gauge theory. From both points of view the theory will confine and the Maxwell term should emerge.

These general arguments are also true in the case of a dimensionless gauge coupling when the theory without a Maxwell term resides at a fixed point/surface at which the scale invariance is not spontaneously broken\footnote{We thank Zohar Komargodski for a discussion on this point.}. An example for such a fixed point is \cite{Banks:1981nn}. If the conformal window starting to open before the fixed end point is accompanied also by a formation of a moduli space such as in the $\phi^6$ vector model studied above, then the scale invariance can be spontaneously broken along the moduli space resulting in the emergence of a Maxwell term. This we have shown to occur in the gauged $\phi^6$ vector theory.

For the 4D, $\mathcal{N}=4$ super conformal SUSY Yang-Mills theory, an  $\mathcal{N}=4$ super symmetrical removal of the Maxwell term does not leave any dynamical terms. For the special gauge theories in $d=5$ we also expect such a term to emerge at a  fixed point which is a part of a moduli space where the scale symmetry can be broken.

In the process of analyzing the various systems we had visited some sideways. It is there where surprises were encountered. We found an additional twist in the 3D $ \phi^6$ theory, a logarithmic confinement by dilatons in addition  to the confining phase induced by the gauge fields when they are massless.  We have found  a Lorentz invariant coset model where the gauged subgroup of the global symmetry was anomalous.

\section*{Acknowledgements}

We thank T.Banks, W.Bardeen, D.Kutasov, R.C.Myers, S.Shenker and in particular O.Aharony and Z.Komargodski for discussions.

The work of E.Rabinovici is partially supported by the American-Israeli Bi-National Science Foundation, a DIP
grant H 52, the Einstein Center at the Hebrew University, the Humbodlt foundation and the Israel
Science Foundation Center of Excellence.

Research at Perimeter Institute is supported by the Government of Canada through Industry Canada and
by the Province of Ontario through the Ministry of Research \& Innovation.

\section*{\Large{Appendices}}

\appendix

\section{Expansion of the bosonic functional determinant in the presence of a gauge field.}
\label{appx:bfncl-det-expan}

In this appendix we expand (\ref{eqn:Aeff-action}) about
$\mathcal{A}_{\mu}=A_{\mu}^AT^A=0$ and verify
(\ref{eqn:quad-Aeff-action}). For simplicity, let us denote
 \bea
 \int_{x}&\equiv&\int d^{d}x \quad,\quad
 \int_{p}\equiv\int {d^{d}p \over (2\pi)^d}
 \quad ,
 \non
 \whG&\equiv&(-\del^2+m^2)^{-1} ~ \Leftrightarrow ~
 G(x-y)=\langle x|\whG|y\rangle=\int_{p} {e^{ip(x-y)} \over p^2+m^2}
 \quad .
 \eea
where $x$ represents the spacetime coordinate and $p$ is the
momentum $d$-vector.

Then one can write
 \begin{multline}
 S_{eff}(\mathcal{A})\equiv\texttt{Tr}\ln \left( -D_{\mu}D^{\mu} +m^{\,2} \right)
 =\texttt{Tr}\ln(\whG^{-1})+\texttt{Tr}\,\whG(\mathcal{A}_{\mu}\mathcal{A}^{\mu}-2i\mathcal{A}_{\mu}\del^{\mu}-i\del^{\mu}\mathcal{A}_{\mu})
 \\
 -{1\over 2} \texttt{Tr}\,\whG(\mathcal{A}_{\mu}\mathcal{A}^{\mu}-2i\mathcal{A}_{\mu}\del^{\mu}-i\del^{\mu}\mathcal{A}_{\mu})
 \whG(\mathcal{A}_{\nu}\mathcal{A}^{\nu}-2i\mathcal{A}_{\nu}\del^{\nu}-i\del^{\nu}\mathcal{A}_{\nu}) + \, . \, . \, .
 \end{multline}
The first term in the above expansion is just a constant and thus
can be discarded from the action. Linear terms in
$\mathcal{A}_{\mu}$ are vanishing since
$\texttt{Tr}\bigl(T^A\bigr)=0$, or alternatively they reveal a total
derivative and thus can be suppressed as well\footnote{This argument
is worthwhile in the case of U(1) where the tracelessness of
generators is not applicable.}. As a result, the expansion starts
from the quadratic terms, which are given by
 \be
 S_{eff}(\mathcal{A})=\texttt{Tr}\,\whG(\mathcal{A}_{\mu}\mathcal{A}^{\mu})
 +{1\over 2} \texttt{Tr}\,\whG(2\mathcal{A}_{\mu}\del^{\mu}+\del^{\mu}\mathcal{A}_{\mu})
 \whG(2\mathcal{A}_{\nu}\del^{\nu}+\del^{\nu}\mathcal{A}_{\nu}) + \, . \, . \, .
 \ee
In what follows we compute this expression term by term, we start
with\footnote{$``\texttt{tr}"$
represents trace over color indices.}
 \be
 I_0\equiv\texttt{Tr}\,\whG(\mathcal{A}_{\mu}\mathcal{A}^{\mu})={\Gamma(1-d/2) \over (4\pi)^{d/2}}
 m^{d-2}~\texttt{tr}\int_{x}\mathcal{A}_{\mu}(x)\mathcal{A}^{\mu}(x) \, ,
 \label{Appxeqn:I_0}
 \ee
where the dimensional regularization formula
 \be
 \int_p \frac{1}{\(p^2 + \Delta \)^n}
 = {1 \over (4\pi)^{d/2}} \, {\Gamma(n-d/2) \over \Gamma(n)}
 \, \Delta^{d/2-n} \, ,
 \label{Appxeqn:dimreg-1}
 \ee
has been used in order to evaluate $G(x,x)$.

Next term we consider is given by
\begin{multline}
 I_1\equiv\texttt{tr}\int_x\int_y\langle x|\,\whG \,(\del^{\mu}\mathcal{A}_{\mu})\,|y\rangle
 \langle y|\,\whG \,(\del^{\nu}\mathcal{A}_{\nu})\,|x\rangle
 =\texttt{tr}\int_x\int_y G^2(x-y) \del^{\mu}\mathcal{A}_{\mu}(x)\del^{\nu}\mathcal{A}_{\nu}(y)
 ~.
\end{multline}
Building on the definition of $\whG$ yields
\begin{multline}
 G^2(x-y)=\int_p\int_q {e^{ip(x-y)} \over [q^2+m^2][(q+p)^2+m^2]}
 ={\Gamma( 2-d/2) \over (4\pi)^{d/2}}\int_p e^{ip(x-y)}p^{\,d-4}
 \\ \times
 \({1\over 4}+{m^2\over p^{\,2} }\)^{d/2-2} \ _{2}F_1\({1\over 2} ~,~ {4-d \over 2} ~,~{3\over 2} ~;~ {1\over 1+4 \, m^2/p^2}\)
 \, .
\end{multline}
To evaluate the above loop integral over momentum $q$ we have used
Feynman parametrization and dimensional regularization. As a result,
we deduce
\begin{multline}
 I_1={\Gamma( 2-d/2) \over (4\pi)^{d/2}}~\texttt{tr}\int_x\int_y \mathcal{A}^{\mu}(x)\mathcal{A}^{\nu}(y)
 \int_p e^{ip(x-y)}p^{\,d-4}p_{\mu}p_{\nu}
 \\ \times
 \({1\over 4}+{m^2\over p^{\,2} }\)^{d/2-2} \ _{2}F_1\({1\over 2} ~,~ {4-d \over 2} ~,~{3\over 2} ~;~ {1\over 1+4 \, m^2/p^2}\)
 \, .
 \label{Appxeqn:I_1}
\end{multline}

Finally, the last two terms which are necessary for the computation
are
\begin{multline}
 I_2\equiv\texttt{tr}\int_x\int_y\langle x|\whG \mathcal{A}_{\mu}\del^{\mu}|y\rangle
 \langle y|\whG \mathcal{A}_{\nu}\del^{\nu}|x\rangle
 \\
 =\texttt{tr}\int_x\int_y\( {\del \over \del y^{\mu}}G(x-y)\mathcal{A}^{\mu}(y)+G(x-y)\del_{\mu}\mathcal{A}^{\mu}(y) \)
 \\ \times
 \( {\del \over \del x^{\nu}}G(y-x)\mathcal{A}^{\nu}(x)+G(y-x)\del_{\nu}\mathcal{A}^{\nu}(x) \)~,
 \label{Appxeqn:I_2}
\end{multline}
and
\begin{multline}
 I_3\equiv\texttt{tr}\int_x\int_y\langle x|\whG \mathcal{A}_{\mu}\del^{\mu}|y\rangle
 \langle y|\,\whG \,(\del^{\nu}\mathcal{A}_{\nu})\,|x\rangle
 \\
 =-\texttt{tr}\int_x\int_y \( {\del \over \del y^{\mu}}G(x-y)\mathcal{A}^{\mu}(y)+G(x-y)\del_{\mu}\mathcal{A}^{\mu}(y) \)
 G(y-x) \del^{\nu}\mathcal{A}_{\nu}(x) \, .
 \label{Appxeqn:I_3}
\end{multline}

With this definitions at hand, the quadratic part of the effective
action for $\mathcal{A}_{\mu}$ can be written as follows
 \be
 S_{eff}(\mathcal{A})=I_0+{I_1 \over 2}+2(I_2+I_3) + \,. \,. \, .
 \label{Appxeqn:AquadrS_eff}
 \ee
Combining (\ref{Appxeqn:I_2}) and (\ref{Appxeqn:I_3}) yields
\begin{multline}
 I_2+I_3=\texttt{tr}\int_x\int_y\( {\del \over \del y^{\mu}}G(x-y)\mathcal{A}^{\mu}(y)+G(x-y)\del_{\mu}\mathcal{A}^{\mu}(y) \){\del \over \del x^{\nu}}G(x-y)\mathcal{A}^{\nu}(x)
 \\
 = -\,\texttt{tr}\int_x \int_y \mathcal{A}^{\mu}(y) \mathcal{A}^{\nu}(x)\int_{p}e^{ip(x-y)}\int_{q} {q_{\mu}q_{\nu} \over [q^2+m^2][(q+p)^2+m^2]}
 \, .
\end{multline}
Using Feynman parametrization, (\ref{Appxeqn:dimreg-1}) and
 \be
 \int_p \frac{p^2}{\(p^2 + \Delta \)^n}
 ={d/2 \over (4\pi)^{d/2}} \, {\Gamma(n-d/2-1) \over \Gamma(n)}
 \, \Delta^{d/2-n+1} \, ,
 \label{Appxeqn:dimreg-2}
 \ee
yields
\begin{multline}
 I_2+I_3= - {\Gamma( 1-d/2) \over (4\pi)^{d/2}} \int_x \int_y \mathcal{A}^{\mu}(y) \mathcal{A}^{\nu}(x)\int_{p}e^{ip(x-y)}
 p^{d-2} \[ \delta_{\mu\nu} \, f_1({m/p})+{p_{\mu}p_{\nu} \over p^2} \, f_2({m/p})\]
 \, ,
 \label{Appxeqn:I_2+I_3}
\end{multline}
where
 \be
 f_1(m/p)={ 1\over 2}
 \({1\over 4}+{m^2\over p^{\,2} }\)^{d/2-1} \ _{2}F_1\({1\over 2} ~,~ {2-d \over 2} ~,~{3\over 2} ~;~ {1\over 1+4 \, m^2/p^2}\)
 \, ,
 \ee
and
\begin{multline}
 f_2(m/p)=-{d-2 \over 24}\({1\over 4}+{m^2\over p^{\,2} }\)^{d/2-2}
 \\\times
 \bigl(3 \, _{2}F_1\[1/2 ~,~ 2-d/2 ~,~3/2 ~;~ (1+4 \, m^2/p^2)^{-1}\]
 \\  +  \
 _{2}F_1\[3/2 ~,~ 2-d/2 ~,~5/2 ~;~ (1+4 \, m^2/p^2)^{-1}\] \bigr)
 \, .
\end{multline}
Substituting now (\ref{Appxeqn:I_0}), (\ref{Appxeqn:I_1}) and
(\ref{Appxeqn:I_2+I_3}) into (\ref{Appxeqn:AquadrS_eff}) we recover
(\ref{eqn:quad-Aeff-action}).

\section{Effective action of the gauged $(\vec\phi^2)^3$ model to leading order in $1/N$.}
\label{appx:eff-action}

In this appendix we derive the effective action of the gauged $(\vec\phi^2)^3$ model \reef{eqn:gauged-scalar-action} to leading order in $1/N$. The derivation is carried out when the system is in the phase with spontaneously broken scale invariance. In particular, the coupling of the dilaton and the $U(1)$ gauge field to the scalar field $\phi$ is elucidated.

By definition, the effective action is given by
 \be
 \Gamma[\vec\phi_{cl}]=W(\vec J, \vec J^{\,\dag})-\int_x\[\vec{J}^{\,\dag}\cdot\vec{\phi }_{cl} + \vec{J}\cdot \vec{\phi}^{\,\dag}_{cl}\]\, ,
 \label{eqn:fulleff}
 \ee
where as usual
 \be
 e^{-W(\vec J, \vec J^\dag)}=Z[\vec J, \vec J^\dag], \quad \vec{\phi }_{cl}(x)={\delta W(\vec J, \vec J^\dag) \over \delta \vec{J}^{\,\dag}(x)}
 , \quad \vec{\phi }_{cl}^{\,\dag}(x)={\delta W(\vec J, \vec J^\dag) \over \delta \vec{J}(x)}
 ~.
 \ee
From  \reef{eqn:partition}, we learn that to leading order in $1/N$
 \bea
 W(\vec J, \vec J^\dag)&=&-\int_x\int_y \vec{J}^{\,\dag}_x G_{xy}\vec{J}_y
 + {1 \over 2}\prod_{i=1}^{i=4}\int_{x_i}\( \vec{J}_{x_2}\cdot{\partial \vec J^{\,\dag}_{x_1} \over \partial x_1^{\mu}}
 -\vec J^{\,\dag}_{x_1}\cdot{\partial \vec{J}_{x_2}\over \partial x_2^{\mu}} \)
 \( \vec{J}_{x_4}\cdot{\partial \vec J^{\,\dag}_{x_3}\over \partial x_3^{\nu}}
 -\vec J^{\,\dag}_{x_3}\cdot{\partial \vec{J}_{x_4}\over \partial x_4^{\nu}}\)
 \non
 &\times&\int_w\int_u G_{x_1,w}G_{w,x_2}D^{\mu\nu}(w-u) G_{x_3,u}G_{u,x_4}
 \non
 &-&{1 \over 2}\prod_{i=1}^{i=4}\int_{x_i}\int_w\int_u \vec J^{\,\dag}_{x_1}\cdot\vec{J}_{x_2} G_{x_1,w}G_{w,x_2}D_{\lambda}(w-u) G_{x_3,u}G_{u,x_4}\vec J^{\,\dag}_{x_3}\cdot\vec{J}_{x_4} +\ldots
 \eea
where the notation of \reef{eqn:dilprop} and \reef{eqn:photonprop} is used to denote propagators of the dilaton and the gauge particle respectively. The subscripts in the above formula indicate the coordinate(s) on which a given quantity depends, while the ellipsis here and thereafter denote higher order terms in the external source and in $1/N$.

Hence,
 \bea
 \vec{\phi }_{cl}(x)=&-&\int_y G_{xy}\vec{J}_y
 -\prod_{i=1}^{i=3}\int_{x_i}\( \vec{J}_{x_2}\cdot{\partial \vec J^{\,\dag}_{x_1} \over \partial x_1^{\mu}}
 -\vec J^{\,\dag}_{x_1}\cdot{\partial \vec{J}_{x_2}\over \partial x_2^{\mu}} \)
 \( \vec{J}_{x_3}{\partial \over \partial x^{\nu}}
 +{\partial \vec{J}_{x_3}\over \partial x_3^{\nu}}\)
 \non
 &\times&\int_w\int_u G_{x_1,w}G_{w,x_2}D^{\mu\nu}(w-u) G_{x_3,u}G_{u,x}
 \non
 &-& \prod_{i=1}^{i=3}\int_{x_i}\int_w\int_u \vec J^{\,\dag}_{x_1}\cdot\vec{J}_{x_2} G_{x_1,w}G_{w,x_2}D_{\lambda}(w-u)  G_{u,x_3}\cdot\vec{J}_{x_3}G_{x,u}+\ldots
 \eea
or equivalently
 \bea
 \vec{J}(x)&=&(\del^2-m^2)\vec{\phi }_{cl}(x)+ \int_w
 \vec{\phi }_{cl}(x)\,D_{\lambda}(x-w) \, \vec{\phi }_{cl}(w)\cdot\vec{\phi}_{cl}^{\,\dag}(w)
 \\
 &+&\int_u\int_w \(\vec\phi_w\cdot{\partial \vec \phi^{\,\dag}_{w} \over \partial w^{\mu}}
 -\vec\phi^{\,\dag}_w\cdot{\partial \vec \phi_{w} \over \partial w^{\mu}}\)
 \(\vec\phi_u{\partial \over \partial x^{\nu}}
 +{\partial \vec \phi_{u} \over \partial u^{\nu}}\)D^{\mu\nu}(w-u) \delta(x-u)
 +\ldots
 \nonumber
 \eea
and similarly for $\vec{\phi }_{cl}^{\,\dag}$ and $\vec J^{\,\dag}$. Substituting back into \reef{eqn:fulleff}, yields
 \bea
 \Gamma[\vec\phi_{cl}]&=&\int_x \, \vec{\phi}^{\,\dag}_{cl}(-\del^2+m^2)\vec{\phi }_{cl}-{1\over 2}\int_u\int_w
 \vec{\phi }_{cl}(u)\cdot\vec{\phi}_{cl}^{\,\dag}(u)\,D_{\lambda}(u-w) \, \vec{\phi }_{cl}(w)\cdot\vec{\phi}_{cl}^{\,\dag}(w)
 \non
 &+&{1\over 2}\int_u\int_w \(\vec\phi_w\cdot{\partial \vec \phi^{\,\dag}_{w} \over \partial w^{\mu}}
 -\vec\phi^{\,\dag}_w\cdot{\partial \vec \phi_{w} \over \partial w^{\mu}}\)
 \(\vec\phi_u{\partial \phi^{\,\dag}_{u}\over \partial x^{\nu}}
 -\phi^{\,\dag}_{u}{\partial \vec \phi_{u} \over \partial u^{\nu}}\)D^{\mu\nu}(u-w)
 +\ldots
 \eea
Using \reef{eqn:photonprop} and reintroducing $A_\mu$ and $i\lambda$, this can be written as follows
 \bea
 \Gamma[\vec\phi_{cl}, i\lambda, A_\mu]&=&\int_x \[ \vec{\phi}^{\,\dag}_{cl}(-D_{\mu}D^{\mu}+m^2)\vec{\phi }_{cl}+i\, \lambda\,\vec{\phi }_{cl}\cdot\vec{\phi}_{cl}^{\,\dag}\]
 \non
 &+& N \int_p   A_{\mu}(-p) A_{\nu}(p)
 \Big[~ {p^2\delta^{\mu\nu}-p^{\mu}p^{\nu} \over 16\pi p}\(  -2{m \over p} + {p^2+4m^2 \over p^2}\arctan {p \over 2m} \)
 \non
 &+&{1 \over 2\al} p^{\mu}p^{\nu}-\theta \, \varepsilon^{\mu\al\nu}p_{\al}~\Big]
 +{1\over 2}\int_u\int_w \, i\lambda(u)\,D^{-1}_{\lambda}(u-w) \,i\lambda(w)
 +\ldots
 \eea
Substituting \reef{eqn:dilprop}, rescaling the fields $i\lambda\rightarrow \sqrt{96\pi m^3/N }\,\psi$, $A_{\mu}\rightarrow\sqrt{24\pi m/N}\,A_{\mu}$ to establish the canonical form of the propagators and redefining the coupling constans $\al\rightarrow24\pi m\,\al$, $\theta\rightarrow \theta/(24\pi m)$
, we obtain the low energy effective action
 \begin{multline}
 \Gamma[\vec\phi_{cl}, \psi, A_\mu]=\int_x \Big[ \vec{\phi}^{\,\dag}_{cl}(-D_{\mu}D^{\mu}+m^2)\vec{\phi }_{cl}
 +{1\over 2}\, \psi\Big(1+{3\over 20}{\del^2\over m^2}+\ldots\Big)(-\del^2\,\psi)
 +{1 \over 4}\,F^{\mu\nu} F_{\mu\nu}
 \\
 +\sqrt{{96\pi m^3\over N}} \,\vec{\phi }_{cl}\cdot\vec{\phi}_{cl}^{\,\dag}\psi
 + {i\theta \over 2} \, \varepsilon^{\al\mu\nu}A_{\al}F_{\mu\nu}+{1 \over 2\al} (\partial_{\mu}A^{\mu})^2 \Big]+\ldots
 \end{multline}
 with $D_{\mu}=\del_{\mu}+i\sqrt{24\pi m\over N}A_{\mu}$.

\section{Expansion of the 3D fermionic functional determinant in the presence of a gauge field.}
\label{appx:ffncl-det-expan}

The aim of this appendix is to compute the contribution of the
fermionic functional determinant
$\texttt{Tr}\log(\displaystyle{\not}\partial-i\displaystyle{\not}V+m)$
to the quadratic part of the effective action
(\ref{eqn:SUSY-eff-action}).

Expanding around $\displaystyle{\not}V=0$ yields
 \begin{multline}
 \texttt{Tr}\log(\displaystyle{\not}\partial-i\displaystyle{\not}V+m)=
 \texttt{Tr}\log(\displaystyle{\not}\partial+m)+
 {1 \over
 2}\texttt{Tr}\[(\displaystyle{\not}\partial+m)^{-1}\displaystyle{\not}V(\displaystyle{\not}\partial+m)^{-1}\displaystyle{\not}V\]
 +\ldots
 \\
 =\texttt{Tr}\log(\displaystyle{\not}\partial+m)-
 \int_p V^{\mu}(-p)V^{\nu}(p)\int_q {2q_{\mu}(p+q)_\nu-2\,m\,q_\al\varepsilon^{\al\mu\nu}-[m^2+q(q+p)]\delta_{\mu\nu}\over[(p+q)^2+m^2][q^2+m^2]}
 +\ldots
 \end{multline}
where we performed a shift in the integration variable $p\rightarrow
p+q$ and used the following identities
 \bea
 \langle x | (\displaystyle{\not}\partial+m)^{-1} | y \rangle
 &=&\int_p {m-i\displaystyle{\not}p \over p^2+m^2}e^{ip(x-y)}~,
 \non
 \texttt{tr}\(
 \gamma^{\mu}_E\gamma^{\nu}_E\gamma^{\al}_E\)&=&-2\,i\,\varepsilon^{\mu\nu\al}
 \quad \(\varepsilon^{012}=1\)~,
 \non
 \texttt{Tr} \displaystyle{\not}a_1\displaystyle{\not}a_2\displaystyle{\not}a_3\displaystyle{\not}a_4
 &=&2[a_1\cdot a_2\,a_3\cdot a_4-a_1\cdot a_3\,a_2\cdot a_4+a_1\cdot a_4\,a_2\cdot a_3]
 \quad .
 \eea

Using Feynman parametrization and the dimensional regularization
scheme (\ref{Appxeqn:dimreg-1}), (\ref{Appxeqn:dimreg-2}),
yields
\begin{multline}
 \texttt{Tr}\log(\displaystyle{\not}\partial-i\displaystyle{\not}V+m)=
 \texttt{Tr}\log(\displaystyle{\not}\partial+m)
 \\
 -\int_p V^{\mu}(-p)V^{\nu}(p)\,{p^2\delta_{\mu\nu}-p_\mu p_\nu \over 16\pi |p|}
 \[ 2{|m|\over |p|}+\(1-4{m^2\over p^2} \) \arctan {|p| \over 2|m|}\]
 \\
 +{m\over 4\pi}\,\varepsilon^{\al\mu\nu}\int_p V_{\mu}(-p)V_{\nu}(p)\,{p_\al\over |p|}\,\arctan {|p| \over 2|m|}
 +\ldots
\end{multline}


\begin{thebibliography}{99}

\bibitem{Kaluza-Klein}
  T.~Kaluza, %Theodor
  ``Zum Unit\"{a}tsproblem in der Physik'',
  Sitzungsber.\ Preuss.\ Akad.\ Wiss.\ Berlin.\ (Math.\ Phys.) {\bf 1921} 966 (1921); \\
  O.~Klein, %Oskar
  ``Quantentheorie und f\"{u}nfdimensionale Relativit\"{a}tstheorie'',
  Zeitschrift f\"{u}r Physik A Hadrons and Nuclei {\bf 37} 12, 895 (1926).
  
%\cite{Bander:1983mg}
\bibitem{Bander:1983mg}
 M.~Bander,
 %``EQUIVALENCE OF LATTICE GAUGE AND SPIN THEORIES,''
 Phys.\ Lett.\  B {\bf 126}, 463 (1983).
 %%CITATION = PHLTA,B126,463;%%

%\cite{Seiberg:1994pq}
\bibitem{Seiberg:1994pq}
  N.~Seiberg,
  %``Electric - magnetic duality in supersymmetric nonAbelian gauge theories,''
  Nucl.\ Phys.\  B {\bf 435}, 129 (1995)
  \href{http://xxx.tau.ac.il/abs/hep-th/9411149}{[arXiv:hep-th/9411149]}.
  %%CITATION = NUPHA,B435,129;%%

%\cite{Suzuki:2010yp}
\bibitem{Suzuki:2010yp}
  M.~Suzuki,
  %``Approximate gauge symmetry of composite vector bosons,''
  Phys.\ Rev.\  D {\bf 82}, 045026 (2010)
  \href{http://xxx.tau.ac.il/abs/1006.1319}{[arXiv:hep-ph/1006.1319]}
  and references therein.
  %%CITATION = PHRVA,D82,045026;%%

%\cite{Komargodski:2010mc}
\bibitem{Komargodski:2010mc}
  Z.~Komargodski,
  %``Vector Mesons and an Interpretation of Seiberg Duality,''
  \href{http://xxx.tau.ac.il/abs/1010.4105}{[arXiv:hep-th/1010.4105]}.
  %%CITATION = ARXIV:1010.4105;%%

%\cite{Dixon:2010gz}
\bibitem{Dixon:2010gz}
  L.~J.~Dixon,
  %``Ultraviolet Behavior of N=8 Supergravity,''
   \href{http://xxx.tau.ac.il/abs/1005.2703}{[arXiv:hep-th/1005.2703]} and references therein.
  %%CITATION = ARXIV:1005.2703;%%

%\cite{Vanhove:2010nf}
\bibitem{Vanhove:2010nf}
  P.~Vanhove,
  %``The critical ultraviolet behaviour of N=8 supergravity amplitudes,''
   \href{http://xxx.tau.ac.il/abs/1004.1392}{[arXiv:hep-th/1004.1392]} and references therein.
  %%CITATION = ARXIV:1004.1392;%%

%\cite{Bossard:2010dq}
\bibitem{Bossard:2010dq}
  G.~Bossard, C.~Hillmann and H.~Nicolai,
  %``E7(7) symmetry in perturbatively quantised N=8 supergravity,''
  JHEP {\bf 1012}, 052 (2010)
   \href{http://xxx.tau.ac.il/abs/1007.5472}{[arXiv:hep-th/1007.5472]} and references therein.
  %%CITATION = JHEPA,1012,052;%%

%\cite{Bardakci:1970nb}
\bibitem{Bardakci:1970nb}
  K.~Bardakci and M.~B.~Halpern,
  %``New dual quark models,''
  Phys.\ Rev.\  D {\bf 3}, 2493 (1971);\\
  %%CITATION = PHRVA,D3,2493;%%
%\cite{Halpern:1971ay}
%\bibitem{Halpern:1971ay}
  M.~B.~Halpern,
  %``The Two faces of a dual pion - quark model,''
  Phys.\ Rev.\  D {\bf 4}, 2398 (1971);\\
  %%CITATION = PHRVA,D4,2398;%%
%\cite{Halpern:1971qj}
%\bibitem{Halpern:1971qj}
  M.~B.~Halpern and C.~B.~Thorn,
  %``Two Faces Of A Dual Pion - Quark Model. 2. Fermions And Other Things,''
  Phys.\ Rev.\  D {\bf 4}, 3084 (1971);\\
  %%CITATION = PHRVA,D4,3084;%%
%\cite{Mandelstam:1973hy}
%\bibitem{Mandelstam:1973hy}
  S.~Mandelstam,
  %``K degeneracy in nonadditive dual resonance models,''
  Phys.\ Rev.\  D {\bf 7}, 3763 (1973);\\
  %%CITATION = PHRVA,D7,3763;%%
%\cite{Mandelstam:1973hu}
%\bibitem{Mandelstam:1973hu}
  S.~Mandelstam,
  %``Simple nonadditive dual resonance model,''
  Phys.\ Rev.\  D {\bf 7}, 3777 (1973).
  %%CITATION = PHRVA,D7,3777;%%

%\cite{Nahm:1987pc}
\bibitem{Nahm:1987pc}
  W.~Nahm,
  Duke Math. J. 54 (1987) 579-613.
  %``QUANTUM FIELD THEORIES IN ONE-DIMENSION AND TWO-DIMENSIONS,''
  %%CITATION = ITP-SB-87-7;%%

%\cite{Bardakci:1987ee}
\bibitem{Bardakci:1987ee}
  K.~Bardakci, E.~Rabinovici and B.~Saering,
  %``String Models With C < 1 Components,''
  Nucl.\ Phys.\  B {\bf 299}, 151 (1988);\\
  %%CITATION = NUPHA,B299,151;%%
  %\cite{Altschuler:1987zb}
  %\bibitem{Altschuler:1987zb}
  D.~Altschuler, K.~Bardakci and E.~Rabinovici,
  %``A CONSTRUCTION OF THE c < 1 MODULAR INVARIANT PARTITION FUNCTIONS,''
  Commun.\ Math.\ Phys.\  {\bf 118}, 241 (1988).
  %%CITATION = CMPHA,118,241;%%

%\cite{Gawedzki:1988hq}
\bibitem{Gawedzki:1988hq}
  K.~Gawedzki and A.~Kupiainen,
  %``G/h Conformal Field Theory from Gauged WZW Model,''
  Phys.\ Lett.\  B {\bf 215}, 119 (1988);
  %%CITATION = PHLTA,B215,119;%%
%\cite{Gawedzki:1988nj}
%\bibitem{Gawedzki:1988nj}
% K.~Gawedzki and A.~Kupiainen,
  %``Coset Construction from Functional Integrals,''
  Nucl.\ Phys.\  B {\bf 320}, 625 (1989).
  %%CITATION = NUPHA,B320,625;%%

%\cite{Eichenherr:1978qa}
\bibitem{Eichenherr:1978qa}
  H.~Eichenherr,
  %``SU(N) Invariant Nonlinear Sigma Models,''
  Nucl.\ Phys.\  B {\bf 146}, 215 (1978)
  [Erratum-ibid.\  B {\bf 155}, 544 (1979)].
  %%CITATION = NUPHA,B146,215;%%

%\cite{D'Adda:1978kp}
\bibitem{D'Adda:1978kp}
  A.~D'Adda, P.~Di Vecchia and M.~Luscher,
  %``Confinement And Chiral Symmetry Breaking In Cp**N-1 Models With Quarks,''
  Nucl.\ Phys.\  B {\bf 152}, 125 (1979).
  %%CITATION = NUPHA,B152,125;%%


%\cite{Witten:1978bc}
\bibitem{Witten:1978bc}
  E.~Witten,
  %``Instantons, The Quark Model, And The 1/N Expansion,''
  Nucl.\ Phys.\  B {\bf 149}, 285 (1979).
  %%CITATION = NUPHA,B149,285;%%

%\cite{Bardeen:1976zh}
\bibitem{Bardeen:1976zh}
  W.~A.~Bardeen, B.~W.~Lee and R.~E.~Shrock,
  %``Phase Transition In The Nonlinear Sigma Model In Two + Epsilon Dimensional
  %Continuum,''
  Phys.\ Rev.\  D {\bf 14}, 985 (1976).
  %%CITATION = PHRVA,D14,985;%%

%\cite{Duane:1979in}
\bibitem{Duane:1979in}
  S.~Duane,
  %``Generalizations Of Cp**N-1 With Nonabelian Gauge Symmetry In (Two +
  %Epsilon) Dimensions,''
  Nucl.\ Phys.\  B {\bf 168}, 32 (1980).
  %%CITATION = NUPHA,B168,32;%%

%\cite{Haber:1980uy}
\bibitem{Haber:1980uy}
  H.~E.~Haber, I.~Hinchliffe and E.~Rabinovici,
  %``The Cp**(N-1) Model With Unconstrained Variables,''
  Nucl.\ Phys.\  B {\bf 172}, 458 (1980).
  %%CITATION = NUPHA,B172,458;%%

%\cite{Bardeen:1983rv}
\bibitem{Bardeen:1983rv}
  W.~A.~Bardeen, M.~Moshe and M.~Bander,
  %``Spontaneous Breaking Of Scale Invariance And The Ultraviolet Fixed Point In
  %$O(N)$ Symmetric ($\phi_3^{\,\,6}$) Theory,''
  Phys.\ Rev.\ Lett.\  {\bf 52}, 1188 (1984),
  %%CITATION = PRLTA,52,1188;%%
  D.~J.~Amit and E.~Rabinovici,
  %``Breaking Of Scale Invariance In $\phi^6$ Theory: Tricriticality And Critical
  %End Points,''
  Nucl.\ Phys.\  B {\bf 257}, 371 (1985).
  %%CITATION = NUPHA,B257,371;%%

 %\cite{Bardeen:1984dx}
\bibitem{Bardeen:1984dx}
  W.~A.~Bardeen, K.~Higashijima and M.~Moshe,
  %``Spontaneous Breaking Of Scale Invariance In A Supersymmetric Model,''
  Nucl.\ Phys.\  B {\bf 250}, 437 (1985);\\
  %%CITATION = NUPHA,B250,437;%%
%\cite{Moshe:2002ra}
%\bibitem{Moshe:2002ra}
  M.~Moshe and J.~Zinn-Justin,
  %``Phase structure of supersymmetric models at finite temperature,''
  Nucl.\ Phys.\  B {\bf 648}, 131 (2003)
  \href{http://xxx.tau.ac.il/abs/hep-th/0209045}{[arXiv:hep-th/0209045]};\\
  %%CITATION = NUPHA,B648,131;%%
%\cite{Feinberg:2005nx}
%\bibitem{Feinberg:2005nx}
  J.~Feinberg, M.~Moshe, M.~Smolkin and J.~Zinn-Justin,
  %``Spontaneous Breaking Of Scale Invariance And Supersymmetric Models At
  %Finite Temperature,''
  Int.\ J.\ Mod.\ Phys.\  A {\bf 20}, 4475 (2005).
  %%CITATION = IMPAE,A20,4475;%%

%\cite{Dunne:1998qy}
\bibitem{Dunne:1998qy}
  G.~V.~Dunne,
  %``Aspects of Chern-Simons theory,''
  \href{http://arxiv.org/abs/hep-th/9902115}{[arXiv:hep-th/9902115]}
  and references therein.
  %%CITATION = HEP-TH/9902115;%%

%\cite{Halliday:1985tg}
\bibitem{Halliday:1985tg}
  I.~G.~Halliday, E.~Rabinovici, A.~Schwimmer and M.~S.~Chanowitz,
  %``Quantization Of Anomalous Two-Dimensional Models,''
  Nucl.\ Phys.\  B {\bf 268}, 413 (1986).
  %%CITATION = NUPHA,B268,413;%%

%\cite{Polyakov:1975rr}
\bibitem{Polyakov:1975rr}
  A.~M.~Polyakov,
  %``Interaction Of Goldstone Particles In Two-Dimensions. Applications To
  %Ferromagnets And Massive Yang-Mills Fields,''
  Phys.\ Lett.\  B {\bf 59}, 79 (1975).
  %%CITATION = PHLTA,B59,79;%%

%\cite{Weinberg:1997rv}
\bibitem{Weinberg:1997rv}
  For the comprehensive reviews on the subject, see i.e.\\
  S.~Weinberg,
  %``Effective field theories in the large N limit,''
  Phys.\ Rev.\  D {\bf 56}, 2303 (1997)
  \href{http://xxx.lanl.gov/abs/hep-th/9706042}{[arXiv:hep-th/9706042]},
  %%CITATION = PHRVA,D56,2303;%%
  \\
  M.~Moshe and J.~Zinn-Justin,
  %``Quantum Field Theory In The Large $N$ Limit: A Review,''
  Phys.\ Rept.\  {\bf 385}, 69 (2003)
  \href{http://xxx.lanl.gov/abs/hep-th/0306133}{[arXiv:hep-th/0306133]}
  \\ and references therein.
  %%CITATION = PRPLC,385,69;%%

%\cite{Rabinovici:1980dn}
\bibitem{Rabinovici:1980dn}
  E.~Rabinovici and S.~Samuel,
  %``The Cp**(N-1) Model: A Strong Coupling Lattice Approach,''
  Phys.\ Lett.\  B {\bf 101}, 323 (1981).
  %%CITATION = PHLTA,B101,323;%%

%\cite{Aharony:2008ug}
\bibitem{Aharony:2008ug}
  O.~Aharony, O.~Bergman, D.~L.~Jafferis and J.~Maldacena,
  %``N=6 superconformal Chern-Simons-matter theories, M2-branes and their
  %gravity duals,''
  JHEP {\bf 0810}, 091 (2008)
   \href{http://xxx.tau.ac.il/abs/0806.1218}{[arXiv:hep-th/0806.1218]}.
  %%CITATION = JHEPA,0810,091;%%

%\cite{Mukhi:2008ux}
\bibitem{Mukhi:2008ux}
  S.~Mukhi and C.~Papageorgakis,
  %``M2 to D2,''
  JHEP {\bf 0805}, 085 (2008)
   \href{http://xxx.tau.ac.il/abs/0803.3218}{[arXiv:hep-th/0803.3218]}.
  %%CITATION = JHEPA,0805,085;%%

%\cite{Klebanov:2002ja}
\bibitem{Klebanov:2002ja}
  I.~R.~Klebanov and A.~M.~Polyakov,
  %``AdS dual of the critical O(N) vector model,''
  Phys.\ Lett.\  B {\bf 550}, 213 (2002)
   \href{http://xxx.lanl.gov/abs/hep-th/0210114}{ [arXiv:hep-th/0210114]}.
  %%CITATION = PHLTA,B550,213;%%

%\cite{Elitzur:2005kz}
\bibitem{Elitzur:2005kz}
  S.~Elitzur, A.~Giveon, M.~Porrati and E.~Rabinovici,
  %``Multitrace deformations of vector and adjoint theories and their
  %holographic duals,''
  JHEP {\bf 0602}, 006 (2006)
   \href{http://xxx.lanl.gov/abs/hep-th/0511061}{[arXiv:hep-th/0511061]}.
  %%CITATION = JHEPA,0602,006;%%

%\cite{Asnin:2009bs}
\bibitem{Asnin:2009bs}
  V.~Asnin, E.~Rabinovici and M.~Smolkin,
  %``On rolling, tunneling and decaying in some large N vector models,''
  JHEP {\bf 0908}, 001 (2009)
  \href{http://xxx.lanl.gov/abs/0905.3526}{[arXiv:hep-th/0905.3526]}.
  %%CITATION = JHEPA,0908,001;%%

%\cite{Park:1994sw}
\bibitem{Park:1994sw}
  S.~H.~Park,
  %``On O(N)-symmetric gauged phi(6)2+1 theory with Chern-Simons term,''
  Phys.\ Rev.\  D {\bf 51}, 5958 (1995)
  \href{http://xxx.lanl.gov/abs/hep-th/9412057}{[arXiv:hep-th/9412057]}.
  %%CITATION = PHRVA,D51,5958;%%

%\cite{Polyakov:1976fu}
\bibitem{Polyakov:1976fu}
  A.~M.~Polyakov,
  %``Quark Confinement And Topology Of Gauge Groups,''
  Nucl.\ Phys.\  B {\bf 120}, 429 (1977).
  %%CITATION = NUPHA,B120,429;%%

%\cite{Gates:1983nr}
\bibitem{Gates:1983nr}
  S.~J.~Gates, M.~T.~Grisaru, M.~Rocek and W.~Siegel,
  %``Superspace, or one thousand and one lessons in supersymmetry,''
  Front.\ Phys.\  {\bf 58}, 1 (1983)
  \href{http://xxx.tau.ac.il/abs/hep-th/0108200}{[arXiv:hep-th/0108200]}.
  %%CITATION = FRPHA,58,1;%%

%\cite{Dirac:1964}
\bibitem{Dirac:1964}
 P.~A.~M.~Dirac, Yeshiva Lectures on Quantum mechanics (Academic Press, New York, 1964 ).

%\cite{Banks:1981nn}
\bibitem{Banks:1981nn}
  T.~Banks and A.~Zaks,
  %``On The Phase Structure Of Vector-Like Gauge Theories With Massless
  %Fermions,''
  Nucl.\ Phys.\  B {\bf 196}, 189 (1982).
  %%CITATION = NUPHA,B196,189;%%










\end{thebibliography}
\end{document}